\documentclass{aa}  

\usepackage{graphicx}
\usepackage{booktabs}
\usepackage[varg]{txfonts}
\usepackage[draft]{hyperref}
\usepackage{natbib}
\bibpunct{(}{)}{;}{a}{}{,}
\begin{document}

   \title{Flares in Open Clusters with K2.}

   \subtitle{I. M45 (Pleiades), M44 (Praesepe) and M67}

   \author{Ekaterina Ilin
          \inst{1}, Sarah J. Schmidt\inst{1},
          James R. A. Davenport\inst{2}
          \and
          Klaus G. Strassmeier\inst{1}
          }

   \institute{Leinbiz Institut für Astrophysik Potsdam\\
              \email{eilin@aip.de}
         \and
            University of Washington
             \email{jrad@uw.edu}
             }

   \date{Received October 9, 2018; accepted December 17, 2018}

 
  \abstract
   {The presence and strength of a stellar magnetic field and activity is rooted in a star’s fundamental parameters such as mass and age. Can flares serve as an accurate stellar "clock"? }
   {To explore if we can quantify an activity-age relation in the form of a flaring-age relation, we measured trends in the flaring rates and energies for stars with different masses and ages.}
   {We investigated the time-domain photometry provided by \textit{Kepler}’s follow-up mission \textit{K2} and searched for flares in three solar metallicity open clusters with well-known ages, M45 (0.125 Gyr), M44 (0.63 Gyr), and M67 (4.3 Gyr). We updated and employed the automated flare finding and analysis pipeline \texttt{Appaloosa}, originally designed for \textit{Kepler}. We introduced a synthetic flare injection and recovery subroutine to ascribe detection and energy recovery rates for flares in a broad energy range for each light curve.}
   {We collected a sample of 1\,761 stars, mostly late-K to mid-M dwarfs and found 751 flare candidates with energies ranging from $4\cdot10^{32}\,$erg to $6\cdot10^{34}\,$erg, of which 596 belong to M45, 155 to M44, and none to M67. We find that flaring activity depends both on $T_\mathrm{eff}$, and age. But all flare frequency distributions have similar slopes with $\alpha \approx2.0-2.4$, supporting a universal flare generation process. We discuss implications for the physical conditions under which flares occur, and how the sample's metallicity and multiplicity affect our results.}
   {}


   \maketitle
%

\section{Introduction}
As stars age, their magnetic activity evolves. A magnetic dynamo should be at work within all stars with outer convection zones~\citep{schatzman_evry_theory_1962}. When the dynamic magnetic field in the interior reaches the stellar surface and interacts with the atmosphere, a variety of phenomena arise, including star spots, chromospheric emission, and flares~\citep{1979cmft.book.....P}. Flares are magnetic reconnection events that lead to a change in field line topology and subsequent energy release~\citep{priest_magnetic_2002}. They reach from the chromosphere to the corona and emit electromagnetic energy in the form of both thermal and non-thermal emission in radio, hard and soft X-rays ($>$ and $\leq 10\,$keV, respectively), UV, and white light~\citep[see][for detailed reviews]{benz_physical_2010, Benz2016}. The amount of energy released during a stellar flare relative to the star's luminosity can exceed the strongest solar flares and even temporarily amplify the optical stellar flux by orders of magnitude for cool dwarfs~\citep{schaefer_superflares_2000, candelaresi_superflare_2014}. Flaring activity's prospective observational availability and strong link to magnetic evolution has incited us to attempt to quantify it as a function of stellar age and mass.
\\
Age, rotation, and magnetic activity are tightly interrelated by the concept of stellar dynamos~\citep{noyes_rotation_1984}. Observational evidence for a relation between stellar rotation, mass, and age ~\citep{radick_activity_1987, patten_evolution_1996, queloz_rotational_1998}, and its theoretical backing~\citep{baliunas_angular_1984, kawaler_angular_1988} gave rise to gyrochronology~\citep{barnes_rotational_2003} $-$ the age-dating of a single main sequence star from its mass and rotation period. \citet{barnes_simple_2010}, \citet{meibom_spin-down_2015} and \citet{barnes_rotation_2016} have found an unambiguous relation to hold for stellar ages from $0.6\,$Gyr (Hyades) up to about $4.3\,$Gyr (M67) for solar type stars. However, this relation varies strongly with stellar mass~\citep{barnes_simple_2010, 2016Natur.529..181V}. Additionally, the detectability of rotation periods in photometric data drops rapidly beyond solar ages, as noted by \citet{2018ApJ...859..167E}.
Secondary magnetic stellar age tracers such as chromospheric activity~\citep{soderblom_chromospheric_1991, pace_chromospheric_2013, lorenzo-oliveira_age-mass-metallicity-activity_2016} or magnetic activity from  Zeeman Doppler Imaging and Zeeman Broadening ~\citep{vidotto_stellar_2014} are being explored but have not yet been developed into full blown age dating techniques.
\\
Among these more or less expedient magnetic age tracers, flaring activity prospectively stands out with respect to practicability. In cool dwarfs, flares do not suffer from a lack of contrast to quiescent luminosity due to their $\sim10\,000\,$K blackbody spectrum~\citep{1992ApJS...78..565H}, and are observable in most common photometric bands~\citep{benz_physical_2010}. 
If flare processes are truly scale invariant~\citep{lacy_uv_1976, shakhovskaya_stellar_1989} and have a sufficiently long cosmic activity timescale, a flaring-age clock may be calibrated for a broad range of stellar masses down to ultra-cool dwarfs~(UCDs, $T_\mathrm{eff}<2\,900$\,K, \citealt{kirkpatrick_dwarfs_1999}) and take over when other age relations break down.
\\
In accordance to dynamo theories, the flaring-age relation is expected to depend on stellar mass. In fact,~\citet{walkowicz_white-light_2011} note that M dwarfs flare more often in white light (but are less energetic) than K dwarfs, and release more energy per time relative to their quiescent flux. However, at least for late-M dwarfs and UCDs, the relation seems weak~\citep{2018ApJ...858...55P}.
\\
On the other hand, some constraints that limit the use of stellar age tracers may apply to a flaring clock as well: If rotation is the main driver of any magnetic activity, a flaring clock cannot be more precise than gyrochronology on an individual star. Short term variations due to stellar activity cycles~\citep{montet_long-term_2017}, breakdown of the relation in a certain age or mass range ~\citep[e.g.][]{pace_chromospheric_2013,2018ApJ...858...55P}, or saturation of flaring activity above some critical Rossby number similar to what is observed in X-ray luminosity~\citep{2011ApJ...743...48W} could diminish the clock's utility. Furthermore, a gyrochronology-type convergence of the flaring-age relation may not occur or occur later than in the case of rotation. Multiplicity of stellar systems affects our implications about the rotation-age relation both drawn from large samples~\citep{douglas_k2_2016, douglas_poking_2017} and regarding individual systems' evolution~\citep{2016AJ....152..113R}. Similar arguments apply to flares. Metallicity influences magnetic activity~\citep{gray_contributions_2006, karoff_influence_2018} and could affect flaring behaviour. A technical constraint is set by a target's total available monitoring time relative to the flare frequency, in particular for ground based observations, but time-resolved photometric surveys like \textit{CoRoT}~\citep{2009A&A...506..411A}, \textit{Kepler}~\citep{jenkins_initial_2010} and the \textit{Transiting Exoplanetary Survey Satellite} (\textit{TESS}; \citealt{ricker_transiting_2014}) have begun to turn the tide.
\\
Evidence for a measurable time scale for flaring activity in low-mass stars has accumulated during the last decade. Based on data from the Sloan Digital Sky Survey ~\citep[SDSS;][]{york_sloan_2000}, both~\citet{kowalski_m_2009} and \citet{hilton_dwarf_2010} found that flaring M0-M6 dwarfs are preferentially found at lower Galactic heights than stars without notable flaring activity but with H$\alpha$ emission, implying a flaring activity lifetime for these dwarfs that is shorter than the chromospheric activity lifetime. \citet{doorsselaere_stellar_2017} used \textit{Kepler} photometry and rotation information from \citet{2014yCat..22110024M} to find that flaring activity is a good predictor of rotation period, which makes a typical flaring activity timescale plausible. Moreover, \citet{clarke_flare_2018} inferred a low intrinsic variability from a largely consistent relation between flaring activity level, rotation and age for pairs of coeval stars with similar masses in a sample of wide binaries found in \textit{Kepler} archives.
\\
In this work, we test if a mass-dependent flaring-age relation exists and can be inferred from flare statistics in open cluster (OC) photometry. OCs observed by \textit{Kepler}'s follow up mission \textit{K2}~\citep{howell_k2_2014} allow us to study large cohorts of coeval stars with a low spread in metallicity. We cover three of these objects, \object{Pleiades}~\citep[M45, $0.125\,$Myr;][]{bell_pre-main-sequence_2012},  \object{Praesepe}~\citep[M44, $0.63\,$Gyr;][]{boudreault_astrometric_2012} and \object{M67}~\citep[$4.3\,$Gyr;][]{dias_fitting_2012}, observed during the mission's campaigns C4 and C5 (Sect. \ref{data}).
\\
We detected flares as bright peaks in \textit{K2} time-domain photometry with the flare finding and analysis pipeline \texttt{Appaloosa}~\citep{davenport_kepler_2016} (Sect. \ref{appaloosa}), using light curves (LCs) de-trended by \citet{aigrain_k2sc:_2016}. Additional validation of flare candidates is obtained from an artificial signal injection and recovery subroutine that characterizes individual LCs with respect to detection rate and energy recovery. We obtain flare frequency distributions (FFDs), determine flare activity indicators and analyse these as functions of $T_\mathrm{eff}$ and age (Sect. \ref{results}). We discuss the role of the targets' multiplicity and metallicity, and put the results in the context of flare physics and coronal heating mechanisms in Sect. \ref{discussion}.
\section{Data}
\label{data}
\textit{K2} time domain photometry is our source of flares, but the archived data need to be cleared of both systematic effects and intrinsic variability before analysis. We collected cluster membership information as well as multiband photometry to derive spectral type and $T_\mathrm{eff}$ for all targets that enabled us to group them by mass and age. We then estimated their luminosities to determine the energies that individual flares release. Each target was observed in $30$\,min cadence for up to $80$\,days yielding LCs in which we searched for flare signatures.
\subsection{De-trended Light Curves}
\begin{figure}
\centering
\includegraphics[width=9cm]{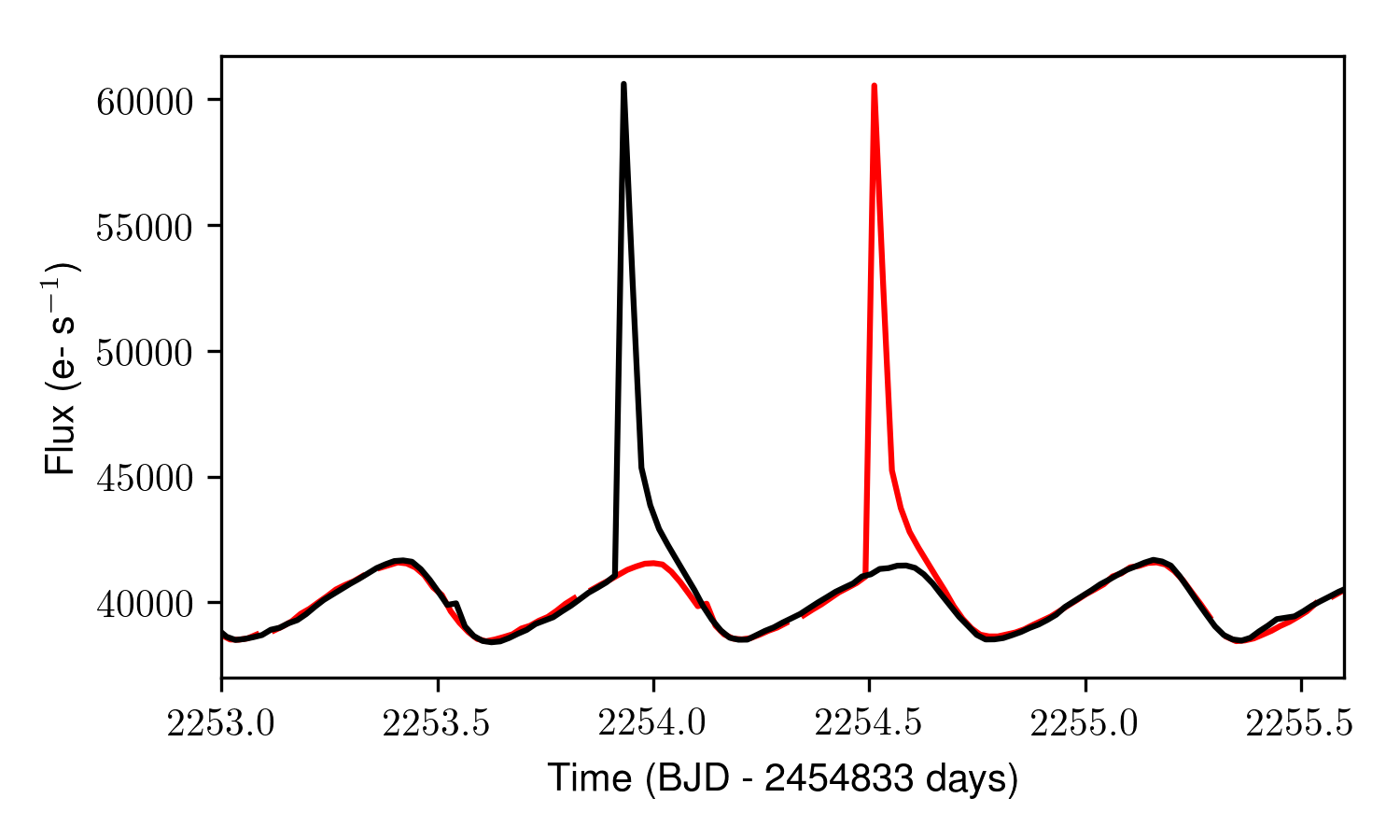}
\caption{$1.8\cdot 10^{34}$\,erg superflare observed on EPIC 211119999 (M45). Black: \textit{K2} PDC\_SAP flux. Red: \texttt{K2SC} residual model with periodic trends added. The \texttt{K2SC} LC is offset by the dominant $0.58$\,d period for visual comparison.}
\label{XM}
\end{figure}

The \textit{Kepler}~\citep{koch_kepler_2010} mission has produced a bounty of high precision photometric observations in the Cygnus-Lyra region since its launch in 2009. In 2013, two reaction wheels failed, and the mission had to be redesigned. In 2014, the follow-up mission \textit{K2}~\citep{howell_k2_2014} has begun to conduct $\sim 80$ days long observation campaigns nearby the ecliptic plane.
Besides the reduced pointing accuracy, the main difference between \textit{Kepler} and \textit{K2} data are additional instrumental artifacts that come along with the challenging task of keeping the satellite in place. Systematic errors occur on different time scales, among which a $6$\,h trend, associated with the spacecraft roll, is most prominent~\citep{van_cleve_thats_2016}. If the roll motion is not properly removed, its shape occasionally resembles flare signatures. 
\\Two de-trending approaches achieve restoration of \textit{Kepler}'s former precision to a very similar degree: the pixel-level de-correlation method developed by \citet{luger_everest:_2016} (\texttt{EVEREST}) and the Gaussian Process (GP) de-trending performed by \citet{aigrain_k2sc:_2016} (\texttt{K2SC}). 
Because their released data products already include LCs with removed periodic signals for campaigns C3$-$C8 and C10, saving considerable computational effort, we opted for \texttt{K2SC} LCs~(see example LC in Fig. \ref{XM}).
\subsection{Open Clusters in K2}
\label{OCK2}
There are about $16$ OCs observed during \textit{K2} campaigns 0$-$18. They span a variety of ages from very young~(M21; $5\,$Myr; \citealt{piskunov_evolution_2011}) to some of the oldest known clusters, such as M67~\citep{howell_k2_2014}. As a rule of thumb, we expect a solar type star to exhibit a flare with $10^{34}$\,erg once every $800$ years~\citep{maehara_superflares_2012}, but, in general, later type stars flare more frequently throughout the whole flare energy spectrum~\citep{doorsselaere_stellar_2017}. By choosing very populous OCs, with $>250$ members each, where every object features \textit{2MASS} and/or \textit{Pan-STARRS} band magnitudes (see Sect. \ref{cat_params} below) as well as de-trended \texttt{K2SC} LCs, we maximize the observation time. Aiming for a broad age range, M45, M44, and M67 were selected for this initial study, so our sample covers ages from $125\,$Myr to $4.3\,$Gyr~(Table \ref{cluster_sample}). 
\paragraph{M45} We use the sample determined by \citet{2016AJ....152..113R}, where membership probabilities are primarily based on recent proper motion studies~\citep{bouy_seven_2015, sarro_cluster_2014, lodieu_astrometric_2012}. All ambiguous candidates were inspected on a case by case basis by \citet{2016AJ....152..113R}. We use the same subset of those candidates except that they exclude the targets without periodicities. This results in a set of high-confidence and lower-confidence members with $6 < K_s < 14.5$, yielding a sample of $826$ stars with \textit{K2} LCs. Lower-confidence means that all proper-motion studies confirmed it was an unambiguous member $-$ as for high-confidence members $-$ but single color-magnitude diagrams (CMDs) placed the star off the respective sequence. We include these, because we later construct our own CMDs and exclude certain targets by a similar criterion.
\paragraph{M44} We use a sample from \citet{kraus_stellar_2007}, who analysed photometry and proper motions of $\sim5$\,million objects to determine their memberships in different stellar populations. \citet{douglas_factory_2014} selected $753$ lower mass ($<1.5M_{\odot}$) and unsaturated ($K_p>9$) stars from this survey with M44 membership and \textit{K2} LCs, and added $41$ known bright members not considered in the survey yielding a total of $794$. We adopt their supplemented selection.
\paragraph{M67} The sample was drawn from \citet{2016MNRAS.459.1060G}, who, similar to \citet{douglas_factory_2014} and \citet{2016AJ....152..113R}, studied stellar variability in \textit{K2} LCs. They started with the sample used in the photometric survey by \citet{nardiello_variable_2016}, which they supplemented with other recent membership studies by \citet{2008A&A...484..609Y} (proper motions) and \citet{geller_stellar_2015} (radial velocities). After checking for consistency in these works they assign, following \citet{geller_stellar_2015}, a membership class to each object from which we only keep those classified as M (members), SM (single members) and BM (binary members), resulting in a working sample of $272$ stars with \textit{K2} LCs.   
\\
\\
From these three samples we removed targets that 
\begin{itemize}
\item lacked multiband photometry and/or empirical template spectra,
\item fell off the main sequence in $g-r$, $r-i$, and/or $J-K$ CMDs,
\item or were assigned spectral types hotter than F4.
\end{itemize}

\begin{table}
\centering
\caption[Cluster sample.]{Cluster sample, partly adopted from \citet{howell_k2_2014}.}
\begin{tabular}{cccc}
\hline\hline
Parameter        & M45 & M44 & M67\\\hline
K2 campaign				 & $4$        & $5$   & $5$  \\
age (Myr)        & $125$      & $630$ & $4\,300$\\
distance (pc)     & $135$      & $160$ & $908$ \\
distance, age (source)     &  (1) & (2) & (3)\\
LCs				 & $737$      & $762$ & $258$ \\
removed targets  & $89$      & $28$  & $14$ \\
$[$Fe/H$]$   		 & $-0.01$    & $0.16$&$0.03$\\
membership       & (4) & (5,6)& (7)\\\hline
\label{cluster_sample}
\end{tabular}
\tablebib{(1)~\citealt{bell_pre-main-sequence_2012};
(2) \citealt{boudreault_astrometric_2012}; (3) \citealt{dias_fitting_2012}; (4) \citealt{2016AJ....152..113R};
(5) \citealt{kraus_stellar_2007}; (6) \citealt{douglas_factory_2014}; (7) \citealt{2016MNRAS.459.1060G}. $[$Fe/H$]$ values are retrieved from \citet{2016A&A...585A.150N}.}
\tablefoot{LCs: number of LCs in the final sample with removed targets subtracted.}
\end{table}

\subsection{Multiband Photometry: $T_\mathrm{eff}$, $R_*$, and Excluded Objects}
\label{cat_params}

To consistently ascribe approximate stellar spectra and ultimately luminosities to the stars we retrieved photometric data from either the \textit{Two Micron All Sky Survey} (\textit{2MASS}; \citealt{skrutskie_two_2006}) or the \textit{Panoramic Survey Telescope and Rapid Response System} \mbox{(\textit{Pan-STARRS})} Data Release 1~\citep{2016arXiv161205560C}. For this, we matched the \textit{K2} sample by position with the \textit{Pan-STARRS} catalog, using the \textit{CDS XMatch} tool\footnote{\url{http://cdsxmatch.u-strasbg.fr/xmatch}} to collect the \textit{Pan-STARRS} $grizy$ measurements (\textit{2MASS} $JHK$ magnitudes are already matched in \textit{EPIC}), which were then converted to SDSS $grizy$ using Table 2 in~\citet{0004-637X-822-2-66}. 
\\
We used the available color indices to derive $T_\mathrm{eff}$ and $R_*$, and eventually luminosities $L_\mathrm{Kp,*}$ that were required to determine individual released flare energies in the \textit{Kepler} band ($E_\mathrm{Kp,flare}$, see Sect. \ref{04}). We employed the synthetic \textit{SDSS}/\textit{2MASS} colors, i.e. $JHK$ and $grizy$ bands, for solar metallicity standards~\citep{pickles_stellar_1998} computed by~\citet{covey_stellar_2007}, and used the \textit{Modern Mean Dwarf Stellar Color and Effective Temperature Sequence}, an up-to-date lookup table for spectral types, $T_\mathrm{eff}$, and $R_*$ derived from $g-r$, $r-i$, $J-H$, and $H-K$ colors, compiled from the literature by \citet{pecaut_intrinsic_2013}\footnote{The updated version used here is 2018.03.22. It includes results from \citet{kirkpatrick_first_2011}, \citet{dieterich_solar_2014-1}, \citet{eker_main-sequence_2015}, \citet{filippazzo_fundamental_2015}, \citet{benedict_solar_2016}, \citet{leggett_near-infrared_2015}, \citet{schneider_hubble_2015}, \citet{patel_sensitive_2014}, \citet{west_sloan_2011}, \citet{dahn_ccd_2017} and \citet{kaltcheva_photometric_2017}.}.
\\
The temperature accuracy $\sigma_{T_\mathrm{eff}}$ in these two tables is heterogeneous for F4 to M9 dwarfs with median $75\,$K and ranging from $20$\,K to $220\,$K~(see Table \ref{classT} for the correspondence of spectral types to $T_\mathrm{eff}$). The lowest accuracies are present for mid-K type stars. The majority of targets had all photometric measurements available. We adopted the median value from typically $3-4$ calculated $T_\mathrm{eff}$. Due to recent findings, e.g. by~\citet{jackson_inflated_2018}, who find that radii for fast rotating M dwarfs in the Pleiades are underpredicted in stellar models by $14\pm2\%$, we assume an uncertainty in $R_*$ of $\sim 20\,\%$. This estimate also covers the spread in observed radii from \citet{pecaut_intrinsic_2013}.
\\
We excluded all likely non-main sequence, early spectral type and foreground/background stars from the sample by examining the CMDs for $g-r$, $r-i$, and $J-K$ colors. The union of objects that fell off the main sequence in these CMDs was rejected. 
\\
The final collection of targets mostly contains late-K to mid-M dwarfs and a few hotter stars. We divided our data in four sufficiently populated bins for stars below $4\,000\,$K and otherwise considered the full sample.
\begin{table}
\centering
\caption{Spectral type and $T_\mathrm{eff}$.}
\begin{tabular}{ccc}
\hline\hline
$T_{min}$ (K) & $T_{max}$ (K) & Spectral types\\\hline
$3\,000$ & $3\,249$ & M3.5$-$M5.5 \\
$3\,250$ & $3\,499$ & M2.5$-$M3 \\
$3\,500$ & $3\,749$ & M1$-$M2 \\
$3\,750$ & $4\,000$ & M0.5$-$K8 \\
$3\,000$ & $7\,000$ & F4$-$M5.5\\\hline
\end{tabular}
\label{classT}
\tablefoot{Correspondence given by~\citet{pecaut_intrinsic_2013}. $T_{min},\,T_{max}$ are the edges of considered $T_\mathrm{eff}$ bins.}
\end{table}
\subsection{Excluded Data}
A subset of data points from all LCs was excluded from further analysis to account for thruster firings and systematics which were not captured by the de-trending pipelines~(see Table \ref{05_flags} for an overview). CR flags were tracked but not removed. 24 severely saturated targets were removed from the sample. After the partial manual review of flare candidates we further excluded some LCs with anomalous variability and individual artifacts.  A detailed description of excluded data and a full list of excluded LCs with high-energy artifacts is given in Appendix \ref{appendix1}.  
\subsection{Flare Energies}
\label{04}
Several individual flare parameters such as duration, amplitude, full width at half the maximum (FWHM), or released energy are typically used in flare statistics~(see e.g.~\citealt{hawley_kepler_2014, yang_flaring_2017, 2018ApJ...859...87Y}). However, in long cadence LCs, duration and amplitude are subject to large errors arising mostly from low time sampling~\citep{2018ApJ...859...87Y}. The flare energy, i.e. the integration of the LC flux during a flare with the quiescent flux subtracted, is less severely affected by this. 
\\
Often, flares are described by a $T_\mathrm{flare}\approx9\,000-10\,000\,$K~\citep{1992ApJS...78..565H, kretzschmar_sun_2011, shibayama_superflares_2013} blackbody Spectral Energy Distribution~(SED). We follow~\citet{shibayama_superflares_2013}, and define the projected stellar quiescent luminosity $L_\mathrm{Kp,*}$ and the flare luminosity in the \textit{Kepler} band $L_\mathrm{Kp,flare}$ as
\begin{equation}
\label{L_*}
L_\mathrm{Kp,*}=\pi R_*^2\displaystyle\int \mathrm{d\lambda}\, R_\mathrm{Kepler}(\lambda)B_{*}(\lambda)
\end{equation}
and
\begin{equation}
\label{Lflare_Kep}
L_\mathrm{Kp,flare}=A_\mathrm{flare}\displaystyle\int \mathrm{d\lambda}\, R_\mathrm{Kepler}(\lambda)B_\mathrm{flare}(\lambda),
\end{equation}
respectively, with $A_\mathrm{flare}$ being the
area  covered  by  the  flaring  region  on  the  stellar  surface, $R_*$ the stellar radius and $R_\mathrm{Kepler}$ the \textit{Kepler} response function given in the \textit{Kepler} instrument handbook~\citep{van2009kepler}. $B_{*}$ and $B_\mathrm{flare}$ correspond to the SEDs of the star and the blackbody curve of the flare with $T_\mathrm{flare}$. The ratio of these luminosities yields the relative flare luminosity $a_\mathrm{Kp,flare}$ as obtained from the \textit{Kepler} LC:
\begin{equation}
a_\mathrm{Kp,flare}=\dfrac{L_\mathrm{Kp,flare}}{L_\mathrm{Kp,*}}
\end{equation}
\citet{kowalski_time-resolved_2013} measured strong variations in $T_\mathrm{flare}$ and flare spectra both between events and during the course of individual flares in several dMe stars, limiting the utility of the blackbody approximation. Additionally, they noted strong time-dependent line emission with a complicated relationship to the continuum. The \textit{Kepler} flare energy $E_\mathrm{Kp,flare}$ is more tractable. Instead of bolometric flare luminosity we use the observed flare luminosity in the \textit{Kepler} band $L_\mathrm{Kp,flare}$ (eqn. \ref{Lflare_Kep}) and obtain
\begin{align}
\label{E_Kep}
E_\mathrm{Kp,flare}&=L_\mathrm{Kp,*}\displaystyle\int_{t_0}^{t_0+\Delta t_\mathrm{flare}} \mathrm{dt}\,a_\mathrm{Kp,flare}\nonumber\\
&=L_\mathrm{Kp,*}\cdot ED.
\end{align}
$ED$ is defined as the area between the LC and the quiescent flux, i.e. the integrated flare flux divided by the median quiescent flux $F_0$ of the star, integrated over the flare duration~\citep{hunt-walker_most_2012}:
\begin{equation}
\label{05_ED}
ED=\displaystyle \int \mathrm dt\, \frac{F_{flare}(t)}{F_0}
\end{equation}
Since it is measured relative to the quiescent star, $ED$ is a quantity independent of calibration and distance. Note, that by this approximation we miss at least the $\sim 27\,\%$ of the continuum flare flux that resides in the U band relative to the total in UBVR~\citep{1992ApJS...78..565H} and flux from emission lines that lie outside the \textit{Kepler} band~(see \citet{kowalski_time-resolved_2013} and references therein). As a consequence, $E_\mathrm{Kp,flare}$ should be considered a lower limit to the total released energy of the flare.
\\
We calculate $L_\mathrm{Kp,*}$ as defined in eqn.~\ref{L_*} using spectra for FGKM stars computed by~\citet{2017ApJ...836...77Y}. Including a parametrized spectrum mitigates errors arising from deviations from the blackbody assumption which becomes particularly relevant as we move to low-mass stars that have strong absorption features in the \textit{Kepler} band along with an ever lower flux overall. \citet{2017ApJ...836...77Y} provide empirical spectra for $3\,000-7\,000$\,K and $0.1-16\,R_\odot$ main sequence and giant branch stars with an approximate accuracy of $\pm100\,$K and $10\,\%$, respectively, rendering them approximately as precise as the color-temperature relations in~\citet{pecaut_intrinsic_2013}. We use Empirical SpecMatch (SpecMatch-Emp\footnote{\url{https://github.com/samuelyeewl/specmatch-emp}}) as a tool to assign template spectra according to the stars' spectral types. The stored spectra have high resolution ($R\approx 60\,000$) and high signal-to-noise ratios ($S/N\approx 150$). The metallicities are centered around solar values with a standard deviation of $\sim0.2$\,dex and maximum deviations of $\pm0.6$\,dex. We include the spectral information from derived $R_*$ and $T_\mathrm{eff}$ to match a spectrum $F_{T_\mathrm{eff},R_*}$ to each star:
\begin{equation}
\label{specL_*}
L_\mathrm{Kp,*}=\pi R_*^2\displaystyle\int \mathrm{d\lambda}\, R_\mathrm{Kepler}F_{T_\mathrm{eff},R_*}(\lambda)B_{*}(\lambda)
\end{equation}
\section{Automated Flare Finding}
\label{appaloosa}
\begin{figure}
\centering
\includegraphics[width=8cm]{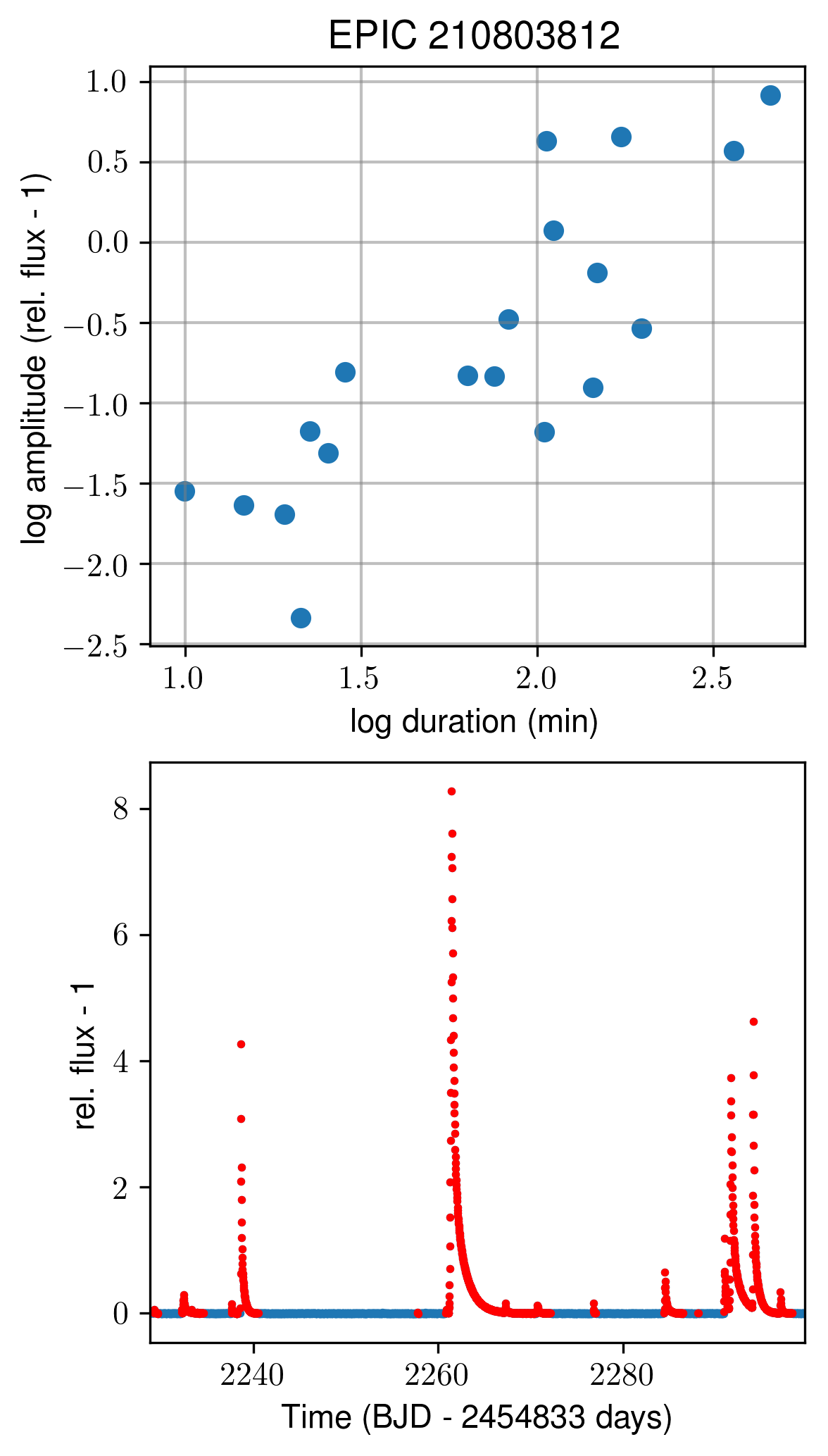}
\caption{Synthetic flare injection for EPIC 210803812 (M45). Left: Distribution of amplitudes and durations of synthetic flare injections. Right: Representation of this distribution in the original \texttt{K2SC} de-trended LC.}
\label{SFI}
\end{figure}

\texttt{Appaloosa} is an open-source\footnote{\url{https://github.com/jradavenport/appaloosa}} flare finding and analysis procedure written in Python by~\citet{davenport_kepler_2016} for \textit{Kepler} LCs. The original version performs two successive steps: First, a model is built for the quiescent stellar brightness. Second, outliers that fulfill a number of detection criteria are analysed and stored in a flare candidate list. 
\\
We skip the first step and employ \texttt{K2SC} LCs, designed by \citet{aigrain_k2sc:_2016} for \textit{K2} data instead, because they already equip us with robust de-trending and variability removal~(see Fig. \ref{XM}). With \texttt{K2SC} LCs, we can treat the residual flux directly and approximate the quiescent flux by the median of all flux measurements.
\\
\citet{chang_photometric_2015} suggest to check every flux outlier for three criteria to determine if it is part of a flare: Outliers from the residual LC are treated as candidates if they exceed thresholds $N_1$ and $N_2$, defined in terms of the LC's variance $\sigma$, i.e.
\begin{align}
\label{chang1}
\dfrac{|f_i-\bar{f}|}{\sigma}&\geq N_1
\end{align}
\begin{align}
\label{chang2}
\dfrac{|f_i-\bar{f}+w_i|}{\sigma}&\geq N_2
\end{align}
where $f_i$ and $w_i$ are the photometric flux and uncertainty at a given time $i$, $\bar{f}$ and $\sigma$ are the median value and the statistical uncertainty in a continuous observation period, as introduced by \citet{chang_photometric_2015} in their \texttt{FINDFlare} algorithm. Eqn. \ref{chang1} and \ref{chang2} define the first two criteria; the third criterion ($N_3$) is the minimum number of consecutive data points that fulfill eqn. \ref{chang1} and \ref{chang2}. In this work, we require at least $3$ consequent data points ($N_3\geq3$, i.e. durations of $\geq 1.5$\,h) for a candidate detection for all outliers that exceed the threshold $N_2=4$. We override eqn.~\ref{chang1} by choosing $N_2>N_1$. For all candidates, the start and end times, and $ED$~(see eqn.~\ref{05_ED}) are extracted. 
\subsection{Flare Finding Efficiency}
\begin{figure}
\centering
\includegraphics[width=8cm]{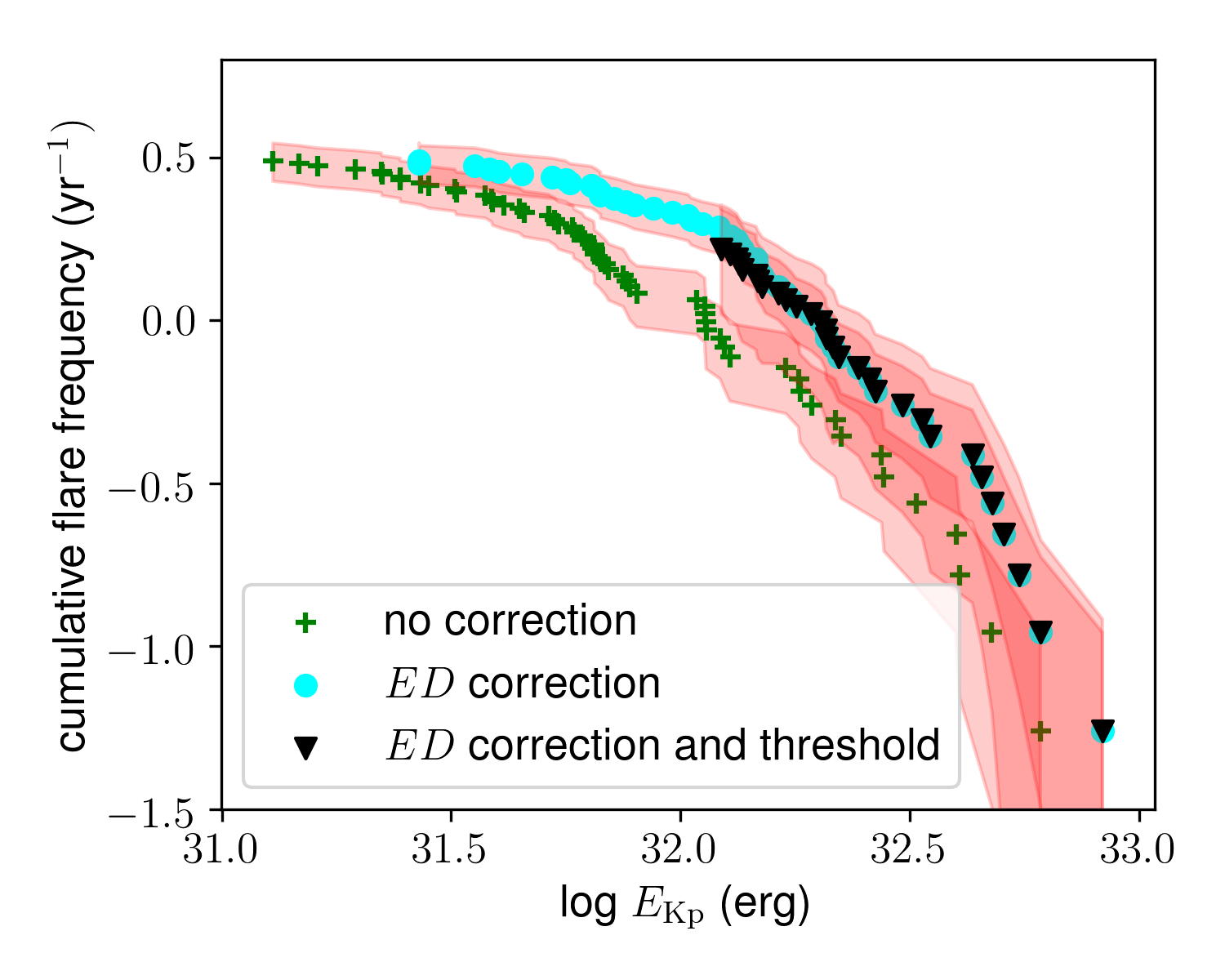}
\caption{Flare recovery and correction scheme for M44 ($3\,500-3\,749$\,K). Green crosses: all detected candidates, no cutoff at low probabilities, no correction for systematic underestimation of $ED$. Blue dots: all detected candidates, no cutoff at low probabilities, but corrected for systematic underestimation of $ED$. Black triangles: Final flare distribution after cutting off at the median detection thresholds for the bin and correcting for systematic underestimation of $ED$. Red shadows: One standard deviation uncertainties.}
\label{SFIexampleFFD}
\end{figure}

A variety of reasons can prevent a flare from being detected in a LC $-$ an event can be lost in the noise, cut off at the end of a continuous observing period, or subjected to filter effects induced by the employed de-trending procedure. Due to the iterative nature of the flare finding procedure and the heterogeneity of LCs it is also not possible to assess the efficiency of the code analytically. We address this problem in a cause-neutral, empirical manner: We test \texttt{Appaloosa}'s flare recovery efficiency by injecting artificial signatures generated from a semi-analytical flare model derived from the active dMe star GJ 1243~\citep{davenport_kepler_2014}. 
\\
The synthetic events are introduced to a LC at different times with varying amplitude and duration while avoiding overlap with real flare signatures~\citep[][see Fig. \ref{SFI}]{davenport_kepler_2016}. The contaminated LC passes through the entire flare finding pipeline. A flare is then considered recovered if the flare peak time is contained within the start and end times of any resulting flare event candidate. After relating all successful and failed detections to each other, the recovery rate as a function of $ED$ is returned. \citet{2018ApJ...858...55P} used this procedure to determine a minimum flare energy detected by the algorithm. We additionally separated this quantity into detection probabilities for individual flare energies and introduced improvements to flare energy recovery and the parameter space covered by the injection routine. Details and examples are given in Appendix \ref{appendix2}.
\\
We specified a scheme for the post-detection treatment of individual flare candidates and the associated uncertainties in cumulative FFDs, illustrated in Fig. \ref{SFIexampleFFD} on the example of M44. 
\\
As a first correction, we adjusted the recorded flare energies according to the energy recovery ratios obtained from synthetic flare injections. This typically shifted the distribution to higher energies in the diagram. In a second step, we rejected all flares with recovery probability below $20\,\%$, a number obtained from experimentation and manual vetting of candidates. We then bolstered the square-root growth of the Poissonian uncertainty by the strongly decreasing detection probability. As $p$ decreases, the count uncertainty grew with $p^{-1/2}$ resulting in correction factors up to $2.2$.
\\
Besides the uncertainties on the event counts and systematic errors on recovered flare $ED$s, we estimated the uncertainties for each flare's $E_\mathrm{Kp,flare}$ (see overview in Table~\ref{06_overview}). \citet{shibayama_superflares_2013} report errors on their energy calculation to be around $\pm 60\%$, which is consistent with our estimate of $\sim 65\,\%$. For most flares, uncertainties mainly stem from the uncertainty on $T_\mathrm{eff}(\text{color})$ and $R_*$, or $L_\mathrm{bol,*}$, which can reach up to $80\,\%$ while the uncertainty on $ED$ is typically below $30\,\%$ with the systematic error accounted for by the aforementioned $ED$ correction. 
\vspace{0.2cm}\\
All in all, the data set's size and quality present a mixed picture. On one hand, the low time resolution of our LCs limits the investigations to high energy (super-)flares. The targets' characterization in terms of $T_\mathrm{eff}$ and $L_\mathrm{bol,*}$ is subject to large uncertainties, not least because of the observed departure of low mass stars from modern stellar models and the lack of standard stars for late-K type stars. These uncertainties propagate all the way through to $E_\mathrm{Kp,flare}$. On the other hand, \texttt{K2SC} de-trended LCs nearly restore the original \textit{Kepler} precision and provide good model fits to the systematic variations of individual LCs while preserving astrophysical signal. Ultimately, our synthetic flare injection procedure allows us to evaluate the quality of each individual LC and additionally validates the detected candidates.
\begin{table}
\caption{Flare energy uncertainties.}
\centering
\begin{tabular}{cccccc}
\hline\hline    
         & LCs & Flares & $\overline{\sigma_{ED}}(\sigma)$ & $\overline{\sigma_{L_{bol}}}(\sigma)$ & $\overline{\sigma_{E_{Kp,flare}}}(\sigma)$\\\hline
M45 & 737 & 596    & 0.29(0.52) & 0.55(0.27) & 0.66(0.52)\\
M44 & 766 & 155    & 0.26(0.24) & 0.53(0.27) & 0.63(0.24)\\
M67	& 258 & 0      & -(-) & -(-) & -(-)\\
\hline
\end{tabular}
\label{06_overview}
\tablefoot{Average uncertainties propagating to flare energies (and their standard deviations). LCs: number of LCs in final analysis. Flares: number of flare candidates used for final analysis.}
\end{table}

\section{Results}
\label{results}
We analyse if flaring activity depends on spectral type and age in late-K to mid-M dwarfs. After a short view on the total flare event counts we introduce activity indicators, i.e., the flaring rate ($FR$), the energy released in flares ($FA$), and the power law fit exponent $\alpha$ and intercept $\beta$ to the flare frequency distributions (FFDs). These indicators complement each other and can be linked to the underpinning astrophysical models as easy to interpret, direct activity measures. Remaining unresolved effects, such as metallicity and multiplicity of the investigated targets, are discussed in Sect. \ref{discussion}.

\subsection{Flare counts}
\label{flare_counts}

We did not find any flare candidates in M67. The two other clusters yielded a final distribution of $751$ flares in total, of which flares in M45 contributed almost four times as many as M44 despite comparable total observation durations. The vast majority of stars and flares were found in the range of late-K to mid-M spectral types. Using these flares, we constructed $5$ FFDs per cluster (Fig. \ref{FFDs}): four for flares on stars with $T_\mathrm{eff}=3\,000-4\,000$\,K divided in $250\,$K bins, and one for the total sample. The choice of bins reflects the balance between the relatively low spectral type resolution, the number of flares per bin and the comparability of bins. Too few flares were detected on stars hotter than $4\,000$\,K to allow for statistical interpretation.
\\
At high energies we expect artifacts to contaminate the tail of the flare distributions. We inspected the LCs and TargetPixelFiles of all flares above $10^{34}$\,erg  using the \texttt{interact} function in \texttt{lightkurve}\footnote{\url{http://doi.org/10.5281/zenodo.1181928}}~\citep{lightkurve}, a dedicated \textit{Kepler} analysis package. As a result, we dropped three targets and 27 individual flares, listed in Table \ref{DropHighEnergy} in Appendix \ref{appendix1}.


\subsection{Flaring Activity Indicators}
\label{FAFR}
The contemporary approach to stellar flaring focuses on empirical studies rather than a robust theoretical description, so that it is unclear which indicator contains the most information about the process' physical underpinning~\citep{Benz2016}. There is no single, generally accepted flaring activity indicator per star or group of stars. We consider four parameters that describe flaring activity and allow us to compare to previous work. The flare rate $FR$ and the energy fraction released in flares $FA$ measure how often flares occur and how strong flaring activity is relative to the quiescent stellar flux. The power law exponent $\alpha$ and intercept $\beta$ can be obtained from fitting a line to the log-log representation of a FFD. $\alpha$ adds differential information to the average values covered by $FA$ und $FR$. $\beta$ depends on strongly on $\alpha$ but it can substitute for $FR$ if $\alpha$ is assumed to be universal. 

\paragraph{FR}
Among statistical flaring activity indicators flare rates are the most directly accessible. By treating a sample of stars in a narrow temperature range as a single prototype star we can define a flaring rate $FR$ as the number of flares $n_i$ recovered from all LCs $i$ added up, divided by the sum of these LCs' observation times $t_i$:
\begin{align}
\label{FR}
FR&=\dfrac{\sum_i n_i}{\sum_i t_i}
\end{align}
$FR$ measures average flare frequency. We treat the uncertainty on $FR$ as Poissonian in the total number of flares.
\paragraph{Flare luminosity}
We define the average total energy released in flares per time as flare luminosity $L_\mathrm{Kp,flare}$ with
\begin{align}
    \label{Lflare}
    L_\mathrm{Kp,flare} &=  \dfrac{\sum_i E_{\mathrm{Kp,flare, tot,} i}}{\sum_i t_i}.
\end{align}
The uncertainty is propagated from the uncertainties on individual $E_\mathrm{Kp,flare}$.
\paragraph{FA}
Alternative to the absolute $L_\mathrm{Kp,flare}$ we introduce a relative activity level measure where the total released energy is related to stellar bolometric luminosity $L_\mathrm{bol,*}$. We consider, adapting the approach in \citet{lurie_kepler_2015}, a flaring activity indicator $FA$ that relates the total flare energy $E_\mathrm{Kp,flare,tot}$ released in the \textit{Kepler} band relative to the estimated bolometric energy release $t_i\cdot L_{\mathrm{bol,*},i}$, where $t_i$ is the observation time of that star:
\begin{align}
\label{FA}
FA&=\dfrac{1}{N}\displaystyle\sum_{i}^{N} FA_i = \dfrac{1}{N} \displaystyle\sum_{i}^{N} \dfrac{E_{\mathrm{Kp,flare,tot,}i}}{t_i\cdot L_{\mathrm{bol,*,}i}}
\end{align}
$N$ is the number of stars in a certain temperature bin. The uncertainty on $FA$ is propagated in quadrature from the uncertainty on $E_{\mathrm{Kp,flare,}i}$ and $L_{\mathrm{bol,*,}i}$. We use bolometric luminosity instead of \textit{Kepler} luminosity because we relate flaring energy to the total energy of a star, but \textit{Kepler} energy contains a different fraction of bolometric energy depending on spectral type. However, we implicitly assume that the fraction of flare energy released in the \textit{Kepler} band does not depend on spectral type, assuming flare production is a universal process in this respect.  
\paragraph{FFD}
\label{results_FFD}

\begin{table*}[ht]
\centering
\caption{FFDs: Results}
\label{alphabeta}
\begin{tabular}{ccccccccc}
\hline\hline
   Cluster &  $T_\mathrm{min}$ &  $T_\mathrm{max}$ &      $n_*$ &  $n_\mathrm{flares}$ &        $\alpha$ &     $\log\beta$ &             $\log\beta_2$ & $\log E_\mathrm{min}$ \\
\hline
       M44 &              3\,000 &              7\,000 &        766 &                  155 &  $2.21\pm0.02$ &  $40.14\pm0.70$ &  $32.99\pm^{0.11}_{0.16}$ &                 32.89 \\
       M44 &              3\,000 &              3\,249 &        206 &                   22 &  $2.05\pm0.02$ &  $34.50\pm0.90$ &  $32.91\pm^{0.28}_{1.31}$ &                 33.05 \\
       M44 &              3\,250 &              3\,499 &        164 &                   55 &  $2.13\pm0.05$ &  $37.50\pm1.79$ &  $33.14\pm^{0.19}_{0.37}$ &                 32.76 \\
       M44 &              3\,500 &              3\,749 &        152 &                   88 &  $2.02\pm0.03$ &  $34.00\pm1.00$ &  $33.28\pm^{0.12}_{0.17}$ &                 32.56 \\
       M44 &              3\,750 &              4\,000 &         47 &                   13 &  $2.01\pm0.07$ &  $33.39\pm2.61$ &  $32.97\pm^{0.20}_{0.38}$ &                 32.60 \\
       M45 &              3\,000 &              7\,000 &        737 &                  596 &  $2.16\pm0.01$ &  $39.03\pm0.53$ &  $33.76\pm^{0.05}_{0.06}$ &                 32.93 \\
       M45 &              3\,000 &              3\,249 &        224 &                   94 &  $2.05\pm0.01$ &  $34.89\pm0.65$ &  $33.18\pm^{0.14}_{0.22}$ &                 32.68 \\
       M45 &              3\,250 &              3\,499 &        195 &                  262 &  $2.14\pm0.02$ &  $38.37\pm0.92$ &  $33.79\pm^{0.08}_{0.10}$ &                 32.77 \\
       M45 &              3\,500 &              3\,749 &        130 &                  179 &  $2.37\pm0.03$ &  $46.13\pm1.21$ &  $33.84\pm^{0.10}_{0.13}$ &                 32.88 \\
       M45 &              3\,750 &              4\,000 &         47 &                   53 &  $2.15\pm0.06$ &  $39.07\pm2.20$ &  $34.04\pm^{0.12}_{0.18}$ &                 32.99 \\
       \hline
\end{tabular}

\tablefoot{Power law parameters to the FFDs as in eqn. \ref{FFDeqn}. $n_*$, $n_\mathrm{flares}$: number of stars and flares in the respective $T_\mathrm{min}-T_\mathrm{max}$ bins. Note that the numbers in $n_\mathrm{flares}$ partly do not sum up, because the median flare energy thresholds in each $T_\mathrm{eff}$ bin vary slightly. The largest bin also overlaps with the $250$\,K bins. $\beta_2$ indicates a least-square fit with $\alpha \equiv 2$. $E_\mathrm{min}$ designates the low-energy detection threshold derived from synthetic flare injection.  }
\end{table*}

\begin{figure}
\centering
\includegraphics[width=7cm]{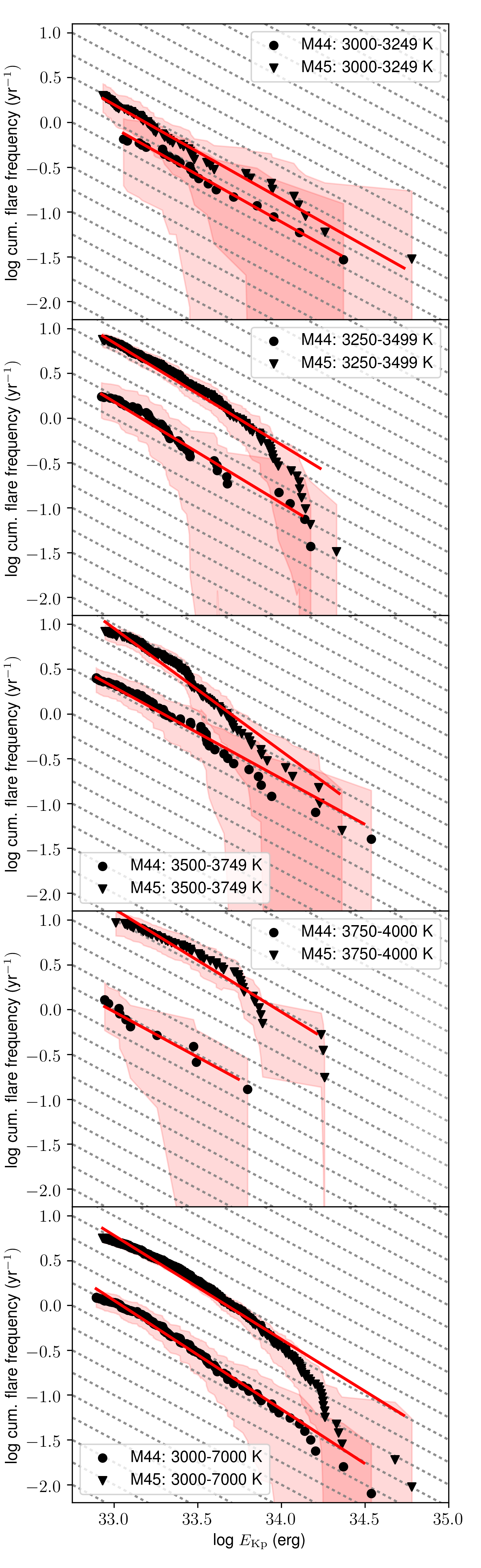}
\caption{Cumulative FFDs. Red lines: ODR fits. Red shadows: weighted Poissonian uncertainties on the frequency. Low energy thresholds are given in Table \ref{alphabeta}. Grey dotted lines indicate a power law with exponent -1 for comparison.}
\label{FFDs}
\end{figure}

\begin{figure}
\centering
\includegraphics[width=8cm]{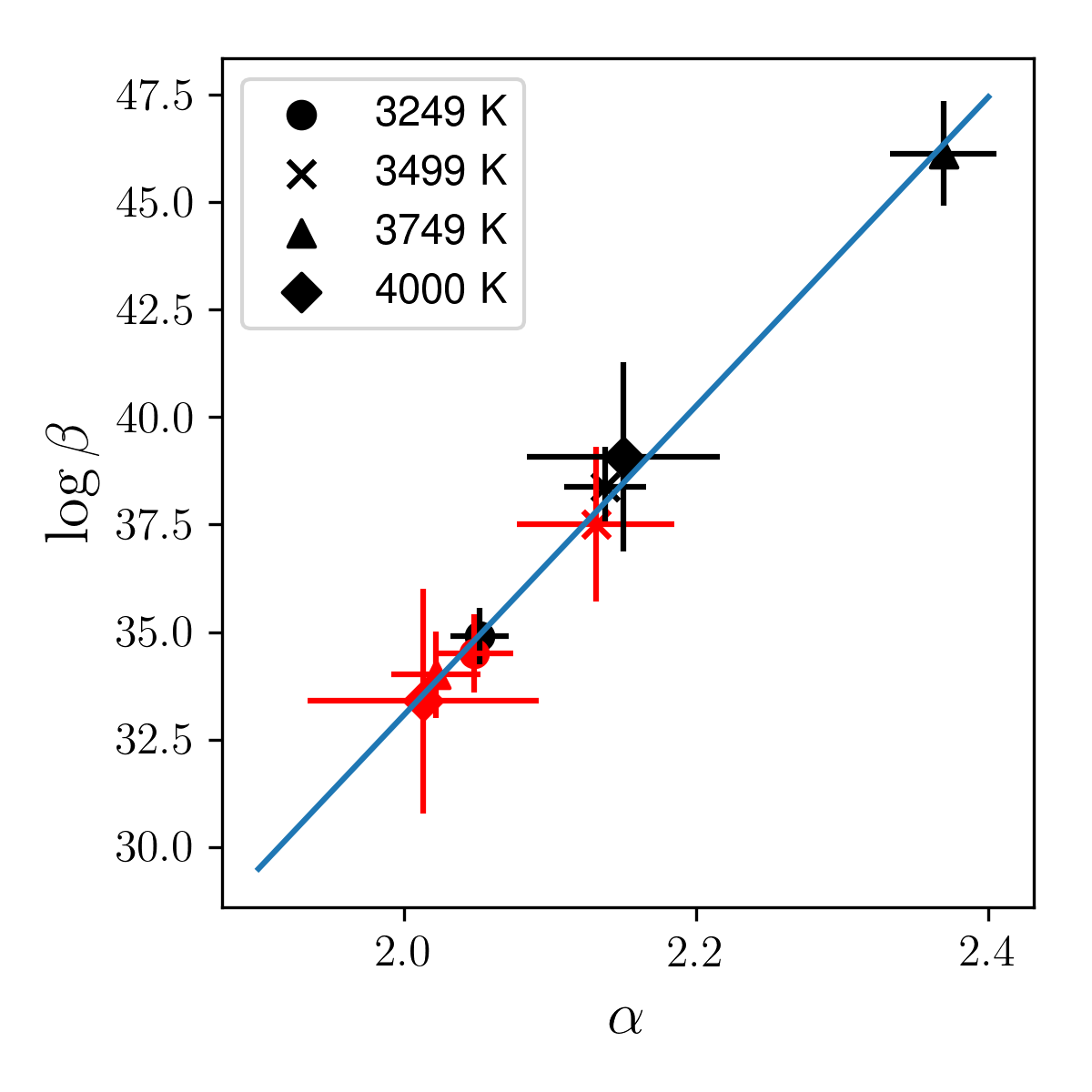}
\caption{Power law parameters to the FFDs, as in eqn. \ref{FFDeqn}. Temperatures in the legend indicate $T_\mathrm{max}$. Black symbols: M45. Red symbols: M44. See also Table \ref{alphabeta}. Blue line: linear least square fit, see eqn. \ref{beta_of_alpha}.}
\label{alpha_vs_beta}
\end{figure}

\begin{figure}
\centering
\includegraphics[width=8cm]{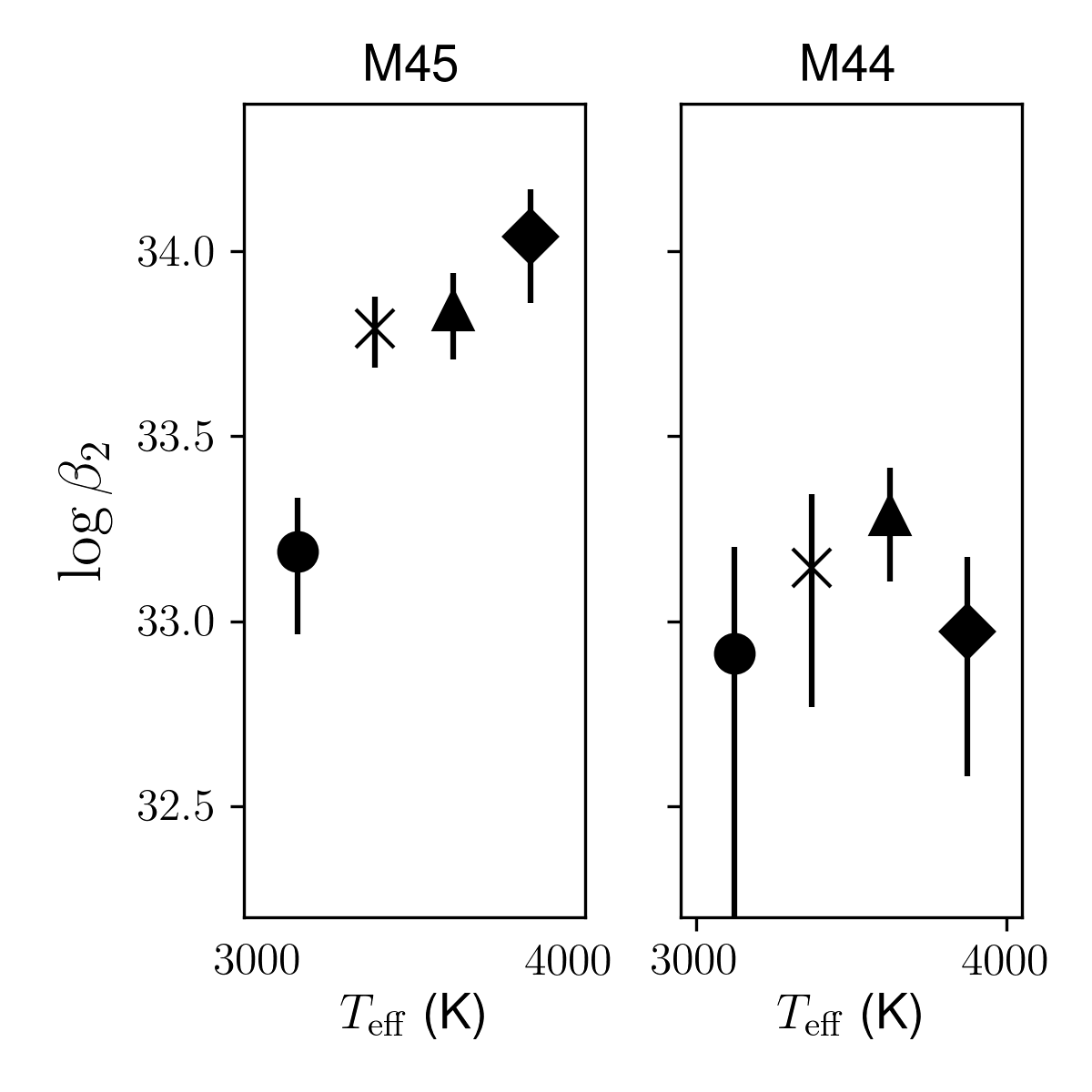}
\caption{Power law parameters to the FFDs, as in eqn. \ref{FFDeqn} with $\alpha\equiv 2$. See also Table \ref{alphabeta}.}
\label{fixedalpha_vs_beta}
\end{figure}

Using our synthetic flare injection procedure, we validated a total of $751$ flare candidates, $596$ in M45, $155$ in M44. In Fig. \ref{FFDs}, we divide the sample into $250$\,K bins and fit power laws to cumulative FFDs of stars with similar spectral type as well as to the entire samples in both clusters. We drop all flare candidates with $E_\mathrm{Kp,flare}$ below the median detection threshold in each bin. The cut accounts for the consequence of superimposing FFDs derived from LCs of different quality.   
\\
The occurrence rate of flares as a function of their energies, also termed flare frequency distribution \citep[FFD;][]{lacy_uv_1976}, can be written as
\begin{equation}
N(E)\,\mathrm{d}E = \beta E^{-\alpha}\,\mathrm{d}E.
\label{FFDeqn}
\end{equation}
In the log-log representation the above relation reads
\begin{equation}
\log N(E)\, \mathrm{d}E = [\log\beta -\alpha\log E]\, \mathrm{d}E.
\label{05_linrel}
\end{equation}
In the corresponding cumulative distribution, a widely used representation in literature~\citep[see][and references therein]{audard_extremeultraviolet_2000, hawley_kepler_2014, 2018ApJ...858...55P}, the exponent becomes $\hat{\alpha} = \alpha -1$. We fit the parameters given in eqn.\,\ref{05_linrel} to the cumulative FFDs in Fig. \ref{FFDs}. We used Orthogonal Distance Regression (ODR) to fit the line because it takes into account both uncertainties on flare rates and energies~ \citep[see][]{2010arXiv1008.4686H}. The method yielded the best fit parameters $\hat{\alpha}$ and $\log \beta$, whose uncertainties we estimated using the delete-1 \textit{jackknife} algorithm~\citep{quenouille1956notes}. 
\\
Table \ref{alphabeta} summarizes the resulting power law fit parameters for all FFDs. Overall, a power law slope $\alpha\approx 2.0-2.4$ is similar in most temperature bins and across both clusters. In the total sample and, marginally, in some of the 250\,K bins, we notice a departure from the single power law at high energies.
\\
The values for $\beta$ are typically dominated by $\alpha$, as we can see in Fig. \ref{alpha_vs_beta}:
\begin{equation}
\log\beta = (35.9\pm1.0)\,\alpha - (38.8\pm 2.1)
\label{beta_of_alpha}
\end{equation}
If we attribute at least some of the deviation from a single power law to pixel saturation, we expect $\alpha$ to be lower. If we then assume $\alpha$ to be universal, we can perform a fit with $\alpha \equiv 2$ and optimize only for the intercept $\log \beta_2$. Fig. \ref{fixedalpha_vs_beta} shows that $\beta_2$ clearly depends on age but less on mass. Overall, $\beta_2$ is consistent with the trends in $FR$.

\subsection{Aging of $FR$, $L_\mathrm{Kp,flare}$, and $FA$}
\begin{figure}
\centering
\includegraphics[width=8cm]{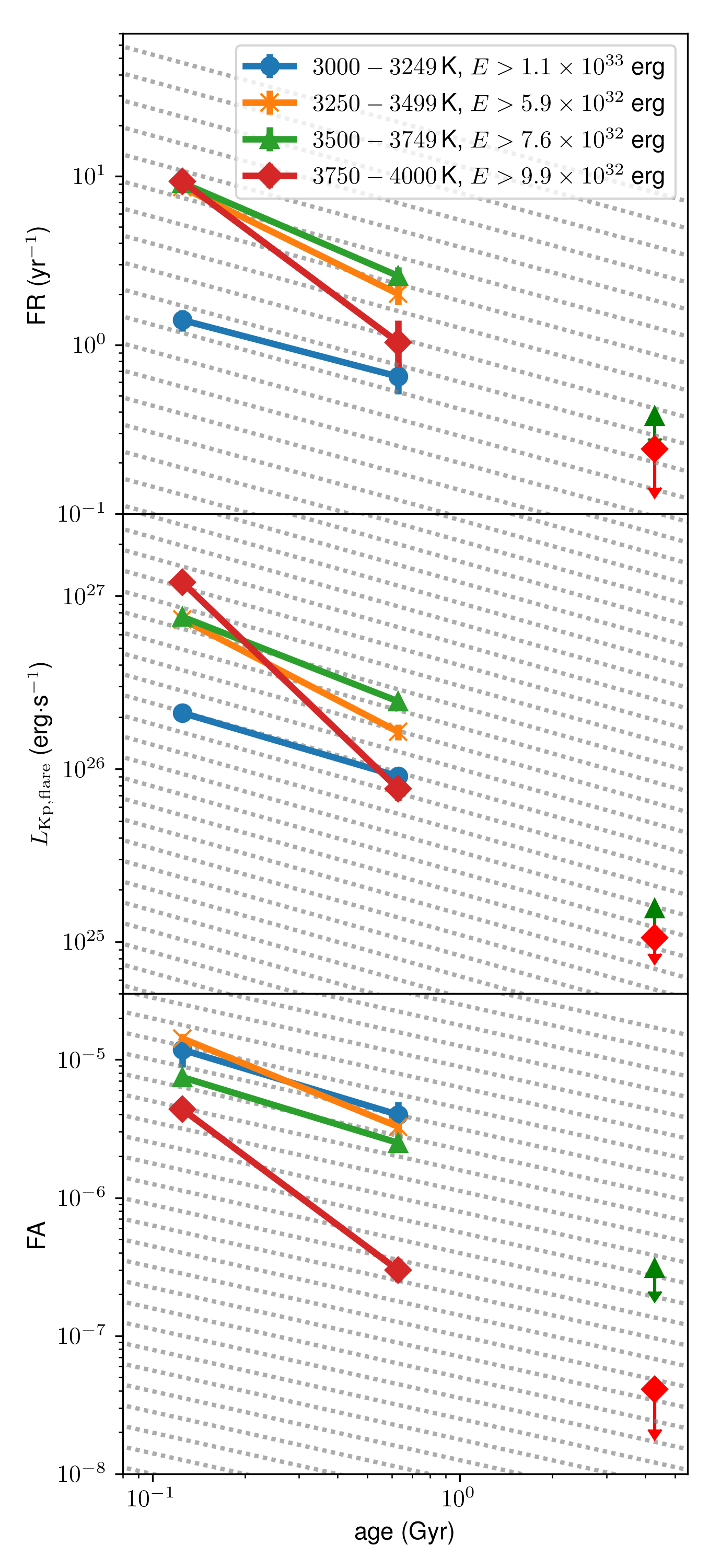}
\caption{Age dependent activity indicators. Color and line shapes distinguish different $T_\mathrm{eff}$ bins. Detection thresholds are derived from synthetic flare injections and averaged in each bin. For M67, available upper limits are given with arrows. The highest energy we detect is $\sim 6\cdot 10^{34}$\,erg. Top: flare rate $FR$; center: absolute flare luminosity in the Kepler band $L_{Kp,flare}$; bottom: Flaring activity $FA$ (see definitions in Sect. \ref{FAFR}). Grey dotted lines indicate the Skumanich age$^{-1/2}$ law for comparison.}
\label{FAFR_age}
\end{figure}
\begin{figure}
\centering
\includegraphics[width=8cm]{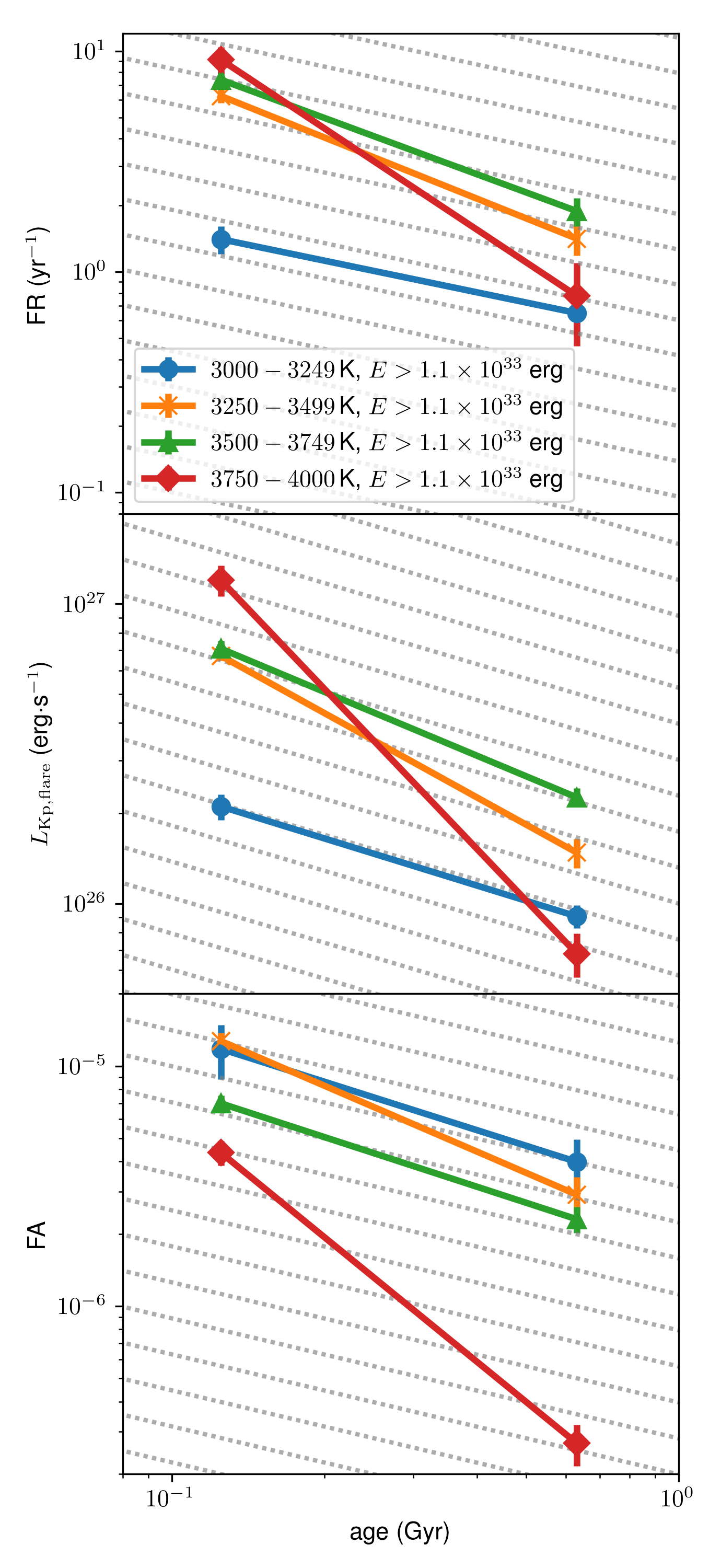}
\caption{Same as Fig. \ref{FAFR_age} but with equal detection thresholds for all $T_\mathrm{eff}$ bins.}
\label{FAFR_age1}
\end{figure}
$FR$, $L_\mathrm{Kp,flare}$, and $FA$ all decline with age (Fig. \ref{FAFR_age}). In the time range from roughly ZAMS to 630\,Myr flaring activity decreases faster with higher $T_\mathrm{eff}$. Stars in the 3\,000-3\,250\,K bin follow the Skumanich $t^{-1/2}$ law relatively closely for both indicators, hotter stars' activity declines much faster. Above the highest detection threshold we can compare activity levels for fixed ages at the cost of higher uncertainties, as shown in Fig. \ref{FAFR_age1}, or alternatively in Fig. \ref{alphaFAFR_T} b.- e.. Fig. \ref{alphaFAFR_T} b. and c. show that $FR$ and $L_\mathrm{Kp,flare}$ are closely correlated. If the distribution from which flare energies are drawn is independent of the flaring rate, i.e., if the flare generation process is universal on all activity levels, $L_\mathrm{Kp,flare}/FR$ will converge for large $FR$. We further discuss the question of universality in Sect. \ref{discussion}.
\\
Our values of $FR$ and $FA$ confirm previous work that suggests an age dependence of flaring activity either from kinematic ages~\citep{kowalski_m_2009, 2011PhDT.......144H}, other OCs like intermediate-age M37 compared to young clusters and the solar neighborhood~\citep{chang_photometric_2015}, or indirectly by showing that rotation, which is an age indicator itself, predicts flaring activity levels~\citep{doorsselaere_stellar_2017}. \citet{clarke_flare_2018} searched \textit{Kepler} time-domain photometry of 33 equal-mass wide binaries for flares. They analysed the relative luminosity emitted in flares on individual stars and found that it was similar within each system: Stars that are alike in mass, metallicity and age exhibit similar magnetic activity levels. As our sample clusters all have similar, close to solar-like metallicities (see Table \ref{cluster_sample}) and all stars in a cluster share approximately the same age, comparing stars with similar masses ($T_\mathrm{eff}$) between clusters should isolate the aging effect on flaring activity within the given uncertainties. 
\\
Recent work by \citet{wright_solar-type_2016} has shown that the saturation of X-ray emission $L_X$ with increasing Rossby numbers $Ro$ (rotation period divided by convective turnover time) is not due to the lack of a tachocline by finding fully convective M dwarfs in the non-saturated regime of the $L_\mathrm{X}(Ro)$ relation. Since flaring activity is tightly correlated with X-ray emission~\citep{neupert_comparison_1968, crosby_frequency_1993, Hannah2011}, our results can be interpreted in the context of the conclusions from \citet{wright_solar-type_2016}: Stars follow a hotter-stars-deplete-faster rule for $FR$, $L_\mathrm{Kp,flare}$, and $FA$ in Fig. \ref{FAFR_age1}. M dwarfs, however, retain high activity levels for several hundred Myr, as \citet{2014AJ....148...64S} found in X-ray, NUV, and FUV observations. 
This rule is consistent with the hotter-stars-spin-down-faster rule derived from stellar rotation studies~\citep{barnes_simple_2010}, and supports the common view that rotation drives magnetic activity~\citep{noyes_rotation_1984}. However, stellar magnetic fields modulate wind driven spin-down~\citep{2018ApJ...862...90G}. We can see that the rotation-activity relationship is tight, but we also know that it is non-linear, both at young~\citep{stauffer_rotation_2016}, and old~\citep{2016Natur.529..181V} ages. For the young and probably fully convective stars, both the apparent saturation in $FA$, and the the lower flaring rates and $L_\mathrm{Kp, flare}$ hint at some regime change, possibly caused by the full convection boundary, and/or the transition from PMS to ZAMS to MS~(see Fig. \ref{FAFR_age1}, blue lines).     
\subsection{Flaring Activity Indicators as a Function of Mass}
\begin{figure}
\centering
\includegraphics[width=7.9cm]{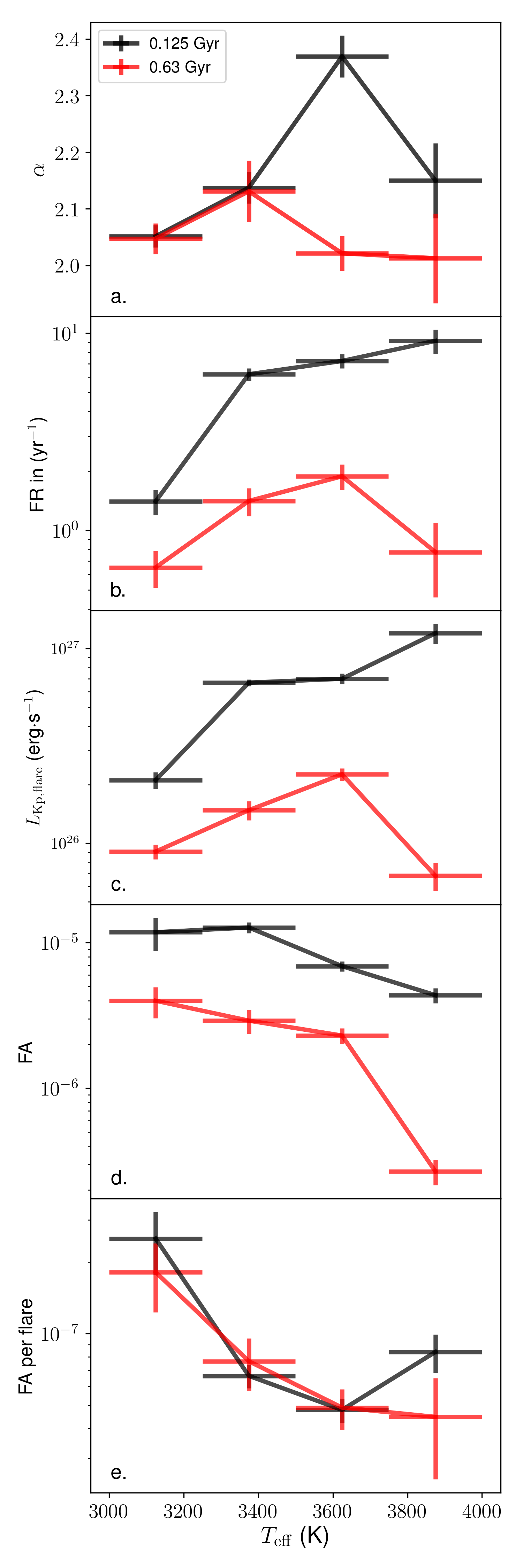}
\caption{Flaring activity measures for M45 and M44 as a function of $T_\mathrm{eff}$. Minimum energy threshold for $L_\mathrm{Kp,flare}$, $FA$, and $FR$: $1.1\cdot 10^{33}$\,erg.}
\label{alphaFAFR_T}
\end{figure}

We can confirm that M dwarfs flare more often than K dwarfs, as \citet{walkowicz_white-light_2011} found in \textit{Kepler} Quarter 1 data~(see also~\citet{candelaresi_superflare_2014} and \citet{doorsselaere_stellar_2017}). We find, for instance, that $FR_{>4000\,\text{K}}/FR_{<4000\,\text{K}}$ is approximately $0.05$ and $0.03$ in M45 and M44, respectively. 
Our flaring rates are seemingly inconsistent with work from \citet{davenport_multi-wavelength_2012}, who found that the time spent flaring increases for mid-M dwarfs compared to early-M type stars~(see Fig. \ref{alphaFAFR_T} c.). The sample in \citet{davenport_multi-wavelength_2012} was mostly composed of field stars so that the young age of our sample could explain the difference. As the authors measure flares in single epochs only, they derive flare luminosities instead of flare energies~(see \citet{davenport_multi-wavelength_2012}, Fig. 7). Assuming typical stellar flare durations of $>10^3$ s, the energy ranges in both works likely overlap in the $\log E_\mathrm{bol,flare} \sim 33.5-34.5$\,erg range. As this is a somewhat narrow range, more detailed analysis may reveal that the distributions are disjoint, which is another possible explanation for the discrepancy.  
\paragraph{FA per flare}
A combination $FA$ and $FR$ is $FA$ \emph{per flare} that measures average flare energy. As opposed to $FA$ and $FR$ individually, $FA$ \emph{per flare} varies with $T_\mathrm{eff}$, but does not exhibit any significant age dependence (see Fig. \ref{alphaFAFR_T} d.). 
\\
In late-M to L type dwarfs, it is believed that the ionization fraction decreases and atmospheric density increases leading to higher resistivity and decoupling of magnetic fields from the atmosphere (relaxation of the frozen-in condition), such that the random walk of surface magnetic loop footprints no longer twists energy into the field topology~\citep{mohanty_activity_2002, rodriguez-barrera_reference_2015}. 
Around $0.3\,M_{\odot}$~($\sim M3.5$, $T_\mathrm{eff}\approx 3\,250\,$K; \citealt{1994sipp.book.....H,2000A&A...364..217D}) stars are thought transition to be fully convective, possibly altering their magnetic topology such that it impacts flare production~(see, e.g.~\citealt{reiners_magnetic_2009}). A $2\,750-3\,000$\,K bin we lack here would more certainly reside below the full convection boundary. These are both possible explanations, but we shall be cautious about interpretations that involve the absolute parameters of our targets.
More generally, different starspot geometries (M. Gully-Santiago, priv. comm.) and varying magnetic field complexities~\citep{2018ApJ...862...90G} need to be considered in a complete picture. 
\section{Discussion}
\label{discussion}
\subsection{Time Evolution of the Power Law Slopes}
We found values of the FFD power law slope $\alpha\approx2.0-2.4$ with no apparent trend with age or mass. The underlying flare production process appears universal on all considered stars. This conclusion is not trivial.
\\
\citet{shakhovskaya_stellar_1989} concluded in her seminal work that flaring activity may be a function of age and that this effect originates in a change in surface magnetic field as the stars spin down over time. She found values for $\alpha$ $\sim 1.6$ in M44 and $\sim 1.8-2.0$ in M45 red dwarfs. More recently, \citet{2018ApJ...858...55P} used kinematic ages of 10 UCDs to discover a weak trend towards shallower $\alpha$ in stellar flaring activity for older ages.
\\
An evolution towards a shallower slope along with an overall decline in $FA$ could be read as follows: Although the overall magnetic energy of a star dissipates over time, the surface magnetic field topology evolves such that longer buildup periods of magnetic stress can occur, caused by slower surface convection and/or rotation. This allows the production of more strong flares relative to weak ones\footnote{The argument works vice versa as well but with shorter buildup periods of magnetic stress that can no longer occur. This allows the production of fewer lower energy flares relative to stronger ones.}. A description of the concrete physical model should involve, besides basic atmospheric parameters like density and temperature with their respective effects on ionization fractions, considerations of mass loss via flare associated cornal mass ejections~\citep{2013ApJ...764..170D, 2018ApJ...862...93A} and field complexity levels on observed bimodal spin-down modes, which are a topic of current research and theoretical studies by themselves~\citep{barnes_simple_2010, newton_rotation_2016-1, 2018ApJ...862...90G}. 
\subsection{Flares as Coronal Heating Mechanisms in late-K to mid-M Dwarfs}
We found $\alpha \gtrapprox 2$ for all temperatures and ages in our results, so it seems possible that flares present the main coronal heating mechanism. 
Since the outset of statistical flare studies, the energy distributions of flares are most frequently described by power laws~\citep{lacy_uv_1976}. The function's exponent, i.e. the slope $\alpha$ in the log-log representation, is highly relevant for our understanding of stellar coronae. Flares have been proposed as coronal heating mechanisms for M dwarfs early by \citet{doyle_ultraviolet_1985}, who found X-ray luminosity to be correlated with the average UV energy released in flares. If $\alpha \geq 2$, the total energy released in flares can be arbitrarily high~\citep{2002ASPC..277..491G}. On the Sun, flare energies can be as low as $10^{26}$\,erg in microflares~\citep{1995PASJ...47..251S}\footnote{We do not consider nanoflares here, because they are produced by a different mechanism than classical flares~\citep{Benz2016}.}. Given a sufficiently steep distribution, one can imagine very low energy flares to be the main coronal heating mechanism. For $\alpha <2$, the highest energy flares contribute most to the total flare energy release and would occur too rarely to be responsible for the quiescent coronal temperature. 
\\
Numbers found in literature all revolve around the value of $2$: \citet{gudel_are_2003} observed AD Leo (dMe3.5) in EUV and soft X-ray and found $\alpha= 2-2.5$, \citet{davenport_multi-wavelength_2012} constrained $\alpha$ to be around $2$ in M0-M6 dwarfs in red-optical and NIR in the \textit{SDSS} and \textit{2MASS} photometric surveys. \citet{lurie_kepler_2015} determined $\alpha\approx 2$ for two dMe5 dwarfs, \citet{gizis_k2_2017-1} found $\alpha=1.8\pm0.2$ for an M8 dwarf, and \citet{2018ApJ...858...55P} studied time domain photometry of $10$ L dwarfs in \textit{K2} data and concluded that $\alpha<2$ for the vast majority of investigated LCs.
\\
Given the proximity of our results to the critical value, and the inconclusive results in other work, the question has to remain open, at least for stars other than the Sun. 
\subsection{Departures from the Power Law Relation in FFDs}
\label{departure}
Flare production is a self-similar process with respect to released energy because we find the FFDs to follow a power law in a broad energy range across different spectral types and stellar ages. A break in or a deviation from this power law would reflect a change in the underlying physics.
\\
A single power law is not always a good fit to our data, as one can see from the full sample FFDs for both clusters in the bottom row in Fig. \ref{FFDs}. Such deviations have already been noticed in single target FFDs, e.g. by \citet{hawley_kepler_2014} and \citet{davenport_kepler_2016}. \citet{2018ApJ...858...55P} find a broken power law and a power law with an exponential cutoff at the highest energies a better fit to the FFDs of two L dwarfs in \textit{K2} LCs. \citet{mullan_frequencies_2018} argue that there could be more than one flare generating regime, i.e. that flares with small energies are fundamentally different from larger flares regarding, e.g., the size of the reconnecting magnetic loop. \citet{2005stam.book.....G} notes that both an instrument's sensitivity/saturation and a maximum energy a star of a certain type can release, may contribute to the deviations they observed for solar type stars.
\\
Assuming that such a maximum energy $E_\mathrm{max, flare}$ exists for late K to mid-M dwarfs, we can argue that the departure from a single power law is the superposition of several stars' FFDs with different $E_\mathrm{max, flare}$. However, in contrast to our FFDs, the deviations discovered in previous work are associated with individual targets' FFDs. \citet{2018ApJ...858...55P}, \citet{hawley_kepler_2014} and  \citet{davenport_kepler_2016} found some but far from all single target FFDs to show this type of departure. An undetected multiplicity, which would create the required superposition of FFDs, offers one explanation to this conspicuity. Another resolution is also plausible: On a single star, multiple active regions may produce a deviation from a single power law. These regions generate flares independently following the same physical process. But each region can have a different $E_\mathrm{max, flare}$. $E_\mathrm{max, flare}$ is limited by a fraction of the maximum magnetic energy available in it~\citep{shibata_can_2013} which in turn depends on the region's size, as~\citet{maehara_starspot_2017} suggest for the Sun and Sun-like stars, and on the regions' geometry as we know from solar observations~\citep{sammis_2000_structure}. A third possibility is that an undetected stellar~\citep{gao_white_2016} or close-in planetary companion~\citep{lanza_star-planet_2012} adds a fundamentally different but morphologically similar flare generation process on top of the intrinsic stellar flare distribution.
\\
However, to test this hypothesis or other physical interpretations, instrumental effects have to be entirely removed first. While our algorithm removed many low-probability flares on the low-energy side, some outliers resulted in a deviation from the fitted power law, which can likely in part be attributed to the heterogeneous detection thresholds determined by the synthetic flare injection procedure. At the high energy end, the deviations can also stem from artifacts and pixel saturation induced systematic errors in the flare energies, some of which we unveiled and removed (see Appendix \ref{appendix1}). As a result, the high-energy end of the distribution affects the slope for the entire sample, which would be shallower if we excluded them. 
\subsection{Metallicity}
We expect metallicity to be a relevant parameter for flare activity studies, because it directly affects the atmosphere within which flaring takes place, but our sample can probably be treated as if metallicity was controled for. 
\\
On one hand, \citet{gray_contributions_2006} found that metal rich stars ($\left[\text{M/H}\right]>-0.2$) have a bimodal chromospheric activity distribution while lower metallicity stars show a single peaked spread.
\citet{karoff_influence_2018} suggested an effect of metallicity on stellar differential rotation and the underlying dynamo\footnote{Here, \citet{karoff_influence_2018} only discuss the classical, tachocline-dependent dynamo paradigm.}. They point out that increasing the metallicity increases the opacity, which in turn will increase the temperature gradient. Then the criterion for convection will be satisfied deeper in the star~\citep{1906WisGo.195...41S}. Deeper convection zones lead to longer convective turnover times near the base of the outer convection zones~\citep{brun_differential_2017}. Stronger differential rotation, a key parameter in the classical $\alpha\Omega$-dynamo, is the consequence~\citep{bessolaz_hunting_2011}.
\\
On the other hand, magnetic activity in wide binaries, that naturally have the same metallicities, is similar within the pairs, as \citet{clarke_flare_2018} show for $33$ equal-mass binaries found in \textit{Kepler} photometry. Thus, the limitation of our OC sample to solar like metallicity clusters dashes joy with pain: We cannot offhand extrapolate the determined age-activity-mass-relation to higher or lower metallicities, but we can mostly exclude that the different activity levels observed in the clusters are an effect of metallicity rather than age: In our sample, M45 has close to solar metallicity, while M44 is even more metal-rich~(see also Table~\ref{cluster_sample}; \citealt{2016A&A...585A.150N}). Unfortunately, the evidence found by \citet{karoff_influence_2018} does not allow us to precisely quantify the effect for M44 because they only compare two solar-type stars. With respect to \citet{gray_contributions_2006} we can neglect the differences in metallicity because all of our studied clusters clearly fall into the metal-rich regime.
\subsection{Multiplicity}
\label{d7}
A quantitative comparison of flaring activity in single and binary members is beyond the scope of this paper, but we note that multiplicity affects our derived activity levels and can cause the misattribution of individual flares to the primary.
\\
Multiplicity is ubiquitous among solar-type and lower mass stars affecting $\sim50\%$ of all systems ~\citep{1991A&A...248..485D,fischer_multiplicity_1992}. \citet{bouvier_formation_2001} observed a binary frequency of about $\sim 50\,\%$, independent of age of the cluster by comparing M44, M45 and the star forming $2\,$Myr old cluster IC 348 for G and K type stars. The frequency is lower for M dwarfs ($42\pm9 \,\%$; \citealt{fischer_multiplicity_1992}) due to the decreasing mass range for companions, and keeps declining towards lower masses, as \citet{boudreault_astrometric_2012} found for the $0.07<M_{\odot}<0.45$ range for M44, but did not fall below $\sim 17\,\%$ within uncertainties for any mass bin they studied. For M67, \citet{geller_stellar_2015} estimated the binary frequency to be as high as $57\pm4\,\%$. Thus, the binary fraction considerably increases the number of individual stars in all clusters for low mass stars, similarly for M44 and M45, and even more notably in M67, where higher mass stars prevail. The effects on both the total activity level and the flare energy distributions are twofold:
\begin{itemize}
    \item The true $FR$ and $FA$ lower, as the true number of stars and hence the cumulated observation time is significantly higher than the number of targets. 
    \item Energies of flares in unresolved binaries are underestimated, regardless of whether they occur on the larger or smaller star, because their $EDs$ are measured relative to the quiescent luminosity of the whole system, but are multiplied only by the luminosity of the large companion that dominates the color indices, while it should be the sum of both~(see eqn. \ref{E_Kep}).
\end{itemize}
The misattribution also implies
\begin{itemize}
    \item a shift of flaring activity levels to cooler stars within the $3\,000-4\,000\,$K range. 
    \item an additional systematic offset to the location of the full convection boundary as may be marked by flares, which would reside at higher assigned spectral types because of the misattribution.
\end{itemize}
We note that the dependence of magnetic activity on the presence of close companions may also affect flaring activity measurements, as discussed by \citet{gao_white_2016}. 
\section{Summary and Conclusions}
Using \texttt{K2SC} de-trended \textit{Kepler/K2} LCs we investigated the flaring activity of three solar metallicity OCs, the ZAMS cluster M45, intermediate age M44 and solar age M67, a total of more than $250$ years of cumulative observation time at $30$\,min cadence from $1\,761$ targets, mostly late-K and early- to mid-M dwarfs. \textit{Pan-STARRS} and \textit{2MASS} multiband photometry yielded $T_\mathrm{eff}$ and radii of individual stars, using solar metallicity standards \citep{pickles_stellar_1998}, computed by \citet{covey_stellar_2007}, and color-temperature relations from \citet{pecaut_intrinsic_2013}. From these we derived quiescent luminosities and the \textit{Kepler} band energies of flares detected by the flare finding and analysis pipeline \texttt{Appaloosa}~\citep{davenport_kepler_2016}.
\\
We improved \texttt{Appaloosa}'s performance using GP Regression de-trended and variablity cleared LCs from \citet{aigrain_k2sc:_2016} instead of raw data. We introduced a synthetic flare injection and recovery routine to characterize long cadence \texttt{K2SC} de-trended LCs of heterogeneous quality. Over $9$ million artificial flare signatures were injected and recovered. They yielded both $ED$-dependent detection thresholds and correction factors to the sampling induced systematic energy underestimation.
\\
We found $751$ flare candidates with $E_\mathrm{Kp,flare}$ ranging from $4\cdot10^{32}\,$erg to $6\cdot10^{34}\,$erg in two clusters, of which $596$ belong to M45 and $155$ to M44. We detected no flare candidates in M67. We saw that both flare rates ($FR$) and energy released relative to bolometric luminosity ($FA$) substantially decline with age for late-K to mid-M dwarfs and follow a hotter-stars-deplete-faster rule. For cooler stars the mass dependence is weak. $FA$ per flare does not show any age dependence and is consistently varying with mass in both clusters. 
\\
Our findings back previous suggestive evidence that a flaring-temperature-age relation exists. Although we do not find an age dependence in the FFD power law exponent $\alpha$, we observe a distinct age dependence for various other activity measures that could prove to be useful age-dating techniques for low-mass stars, complementing existing methods. The lack of any trend in $\alpha$ suggests the universality of the flare production process across a broad range of masses, presumably below the full convection boundary. If $\alpha$ is universal, the power law intercept $\beta_2$ can serve as another age-dependent activity measure. 
\\
Our results are at least valid for stars with solar or close to solar metallicities. We acknowledge that unresolved multiplicity causes misattribution of flares to the primaries, and an overestimation of flaring activity overall. 
\\
Assuming that our FFDs can be described by single power laws for all flare energies, it remains unclear if flares can be the main coronal heating mechanism because the power law exponents $\alpha$ are close to and above the critical value of $2$. A single power law is not always a good fit to the FFDs, although it describes the distributions in most individual temperature bins well. Pixel saturation effects at the FFDs' high-energy ends can cause such deviations. Alternatively, a high energy flaring limit that varies among stars could produce the broken power law by superimposing their FFDs.
\\
In addition to M45, M44 and M67, $K2$ has endowed us with months of continuous monitoring of several OCs, spanning a range of ages from PMS to solar age. We will expand the analysis described in this paper to these targets in an upcoming project, and explore the effects of multiplicity and metallicity on the gears of a stellar flaring "clock".
\\
\citet{macdonald_magnetic_2013, feiden_magnetic_2013}, and many others are working to pin down how magnetic activity affects modern stellar evolution. A comprehensive description of flaring activity as a function of age, which we attempted to approach here, may be one of the bottlenecks in our evolution models of exoplanetary atmospheres~\citep{johnstone_evolution_2015}.
\begin{acknowledgements}
We are grateful to the anonymous referee for their constructive suggestions that significantly improved our interpretation of the results. We also wish to thank Sydney Barnes, Meetu Verma, and Jens Fischer for helpful comments and support.
\\
This research made use of pandas~\citep{mckinney}; Astropy, a community-developed core Python package for Astronomy~\citep{2013A&A...558A..33A}; NumPy \citep{van2011numpy}; and matplotlib, a Python library for publication quality graphics \citep{Hunter:2007}. This research also used the cross-match service provided by \textit{CDS}, Strasbourg.
\\
Some/all of the data presented in this paper were obtained from the Mikulski Archive for Space Telescopes (MAST). STScI is operated by the Association of Universities for Research in Astronomy, Inc., under NASA contract NAS5-26555. This paper includes data collected by the \textit{Kepler/K2} mission. Funding for the Kepler mission is provided by the NASA Science Mission directorate. We made use of data products from the \textit{Two Micron All Sky Survey}, which is a joint project of the University of Massachusetts and the Infrared Processing and Analysis Center/California Institute of Technology, funded by the National Aeronautics and Space Administration and the National Science Foundation. The \textit{Pan-STARRS1} Surveys (\textit{PS1}) and the \textit{PS1} public science archive have been made possible through contributions by the Institute for Astronomy of the University of Hawaii, the \textit{Pan-STARRS} Project Office, the Max-Planck Society and its participating institutes, the Max Planck Institute for Astronomy, Heidelberg and the Max Planck Institute for Extraterrestrial Physics, Garching, The Johns Hopkins University, Durham University, the University of Edinburgh, the Queen's University Belfast, the Harvard-Smithsonian Center for Astrophysics, the Las Cumbres Observatory Global Telescope Network Incorporated, the National Central University of Taiwan, the Space Telescope Science Institute, the National Aeronautics and Space Administration under Grant No. NNX08AR22G issued through the Planetary Science Division of the NASA Science Mission Directorate, the National Science Foundation Grant No. AST-1238877, the University of Maryland, Eotvos Lorand University (ELTE), the Los Alamos National Laboratory, and the Gordon and Betty Moore Foundation.
\\
JRAD is supported by an NSF Astronomy and Astrophysics Postdoctoral Fellowship under award AST-1501418.

\end{acknowledgements}


\bibliography{MyLibrary}


\begin{appendix} 
\section{Excluded Data}
\label{appendix1}

\paragraph{Cosmic Ray Contamination}
Cosmic ray (CR) events occur randomly, possibly adding an offset to the flaring rate of a star because they may be mimicking flare signatures. Unlike the original \textit{Kepler} mission, \textit{K2} does not identify and remove cosmic ray signals in optimal aperture due to the reduced calibration and photometric analysis pipelines. CRs in pixels collateral to optimal aperture, however, are flagged and can be used to extract information about the data quality. In Campaigns 0-11 a CR is flagged if the counts in a pixel exceed the $4\sigma$ limit compared to a rolling band median of five cadences. In our sample, CR flags make up $\sim4-7\,\%$ of all data points depending on the campaign. 
Note that from Campaign 11 on the threshold was increased to $7\sigma$ to minimize false detections triggered by \textit{K2}'s roll motion\footnote{\url{https://keplerscience.arc.nasa.gov/k2-pipeline-release-notes.html\#data-release-11}, 13.03.2018}.
\\
A CR hitting the optimal aperture or spilling signal into it from collateral pixels and signals stemming from flares are possibly ambiguous in long cadence data. Both produce a sudden increase in flux. A higher temporal resolution mitigates the problem: True flares can be identified by the characteristic shape of the flare signal with its impulsive rise and decline followed by a more gradual decay phase. For flares with durations shorter than $\sim1.5\,$h, or $3$ data points, \textit{K2} long cadence data do not resolve any of these features and we do not include any candidates with less than $3$ consecutive data points in our analysis (i.e. $N_3 \geq 3$, see Sect. \ref{appaloosa}). 
\\
 A CR induced flaring rate offset can vary, particularly when the spacecraft drifts and changes its attitude towards the Sun until it rotates again for a new campaign. Different campaigns and the relative position of the target on the CCD can play a role: The spacecraft may shield certain parts of the detector from CR impact better than others.
\\
In this work, we acknowledge the possible CR contamination in our data but treat them as a statistical offset that affects all LCs in a similar manner: We interpret the similar amounts of CR flags we found in all OC samples (see Table \ref{05_flags}) to indicate comparability of the OCs. The weak trend may even reflect true flaring activity differences in the sample. 
\paragraph{Thruster Firings and Field-Wide Systematics}
\label{thruster}
We removed data points labeled as coinciding with the spacecraft's thruster firings as well as their preceding and succeeding measurements that are often flagged as CRs.  This procedure excludes $11-18\,\%$ of all data points from the individual LCs~(see Table \ref{05_flags}). 
\\
Even if the de-trending procedure removes most systematic effects from the LC, some may remain undetected. To identify false positive detections, we compare all LCs and their respective flare candidate time to each other in order to find field-wide systematics. We assume that flares occur randomly and reject as precautionary measure all flare detections that are detected in more LCs simultaneously than this assumption would allow. 
\paragraph{Saturated targets and artifacts}
According to~\citet{van2009kepler}, if the flux in a pixel exceeds the full well depth of 10093\,DN by a factor $<10$, the detector response may still be linear if one considers the total flux in a sufficiently large aperture. We excluded 24 targets from our analysis, where the detected flare candidates exceed this threshold, listed in Table \ref{saturated}. Finally, we vetted the $\sim100$ most energetic flares by eye and excluded uncertain, clearly artificial signals or failed LCs from the results~(see Table \ref{DropHighEnergy}).
\begin{table}[ht]
\caption{Excluded data: Flagged data points.}
\centering
\begin{tabular}{ccccc}
\hline\hline
Cluster & C & $\%$ thruster firings & $\overline{\% \text{CR}}$ & Field-wide\\\hline
M45 & $4$ & $3.85$ & $6.8\pm2.75$ & $9\,(33)$\\
M44 & $5$ & $4.53$ & $5.2\pm 3.24$ & $8\,(28)$ \\
M67 & $5$ & $4.53$ & $4.0\pm 1.60$ & $0\,(-)$\\
\end{tabular}
\label{05_flags}
\tablefoot{Both possible and confirmed thruster firings were removed from the data. C: campaign. $\%$ thruster firings: percentage of data points flagged as (possible) thruster firings. $\overline{\% \text{CR}}$: average contamination of LCs with CR flags.  Field-wide: regions excluded due to simultaneous flares (total number of data points they comprise).}
\end{table}
\begin{table}[ht]
\caption{Excluded data: Extremely saturated targets.}
\centering
\label{saturated}
\begin{tabular}{cc}
\hline\hline    
Cluster & EPIC ID\\\hline
M45 & 210877423 \\
    & 210879932 \\
    & 210966700 \\
    & 210978650 \\
    & 210997197 \\ 
    & 211014186 \\ 
    & 211020371 \\ 
    & 211033155 \\ 
    & 211053737 \\ 
    & 211063235 \\ 
    & 211066337 \\ 
    & 211067634 \\ 
    & 211072441 \\ 
    & 211086025 \\ 
    & 211087059 \\ 
    & 211093684 \\ 
    & 211096368 \\ 
    & 211101761 \\ 
    & 211115638 \\ 
    & 211120842 \\ 
    & 211125210 \\ 
    & 211132233 \\
M44 & 211934056\\
    & 212034371\\
    \hline
\end{tabular}
\end{table}

\begin{table*}[ht]
\caption{Excluded data: Rejected after manual inspection}
\centering
\label{DropHighEnergy}
\begin{tabular}{cccc}
\hline\hline
      EPIC &        $t_0$ &        $t_f$ &                    Note \\
\hline
 210966700 &  2258.671619 &  2258.732914 &        flagged cadences \\
 211038389 &  2251.643200 &  2251.867949 & entire aperture flashes \\
 211038389 &  2252.051835 &  2252.215290 &         no flares in LC \\
 211060530 &              &              &          bright CCD row \\
 211066337 &  2279.471032 &  2279.532327 &   no clear brightening  \\
 211066337 &  2236.993468 &  2237.034332 &   no clear brightening  \\
 211073598 &	 	  &		 &  detector edge	   \\
 211089323 &              &              &               saturated \\
 211091848 &  2243.245708 &  2243.286571 &   no clear brightening  \\
 211096368 &  2274.996528 &  2275.078254 &          bright CCD row \\
 211114329 &  2271.094166 &  2271.135030 &   no clear brightening  \\
 211120842 &  2243.654339 &  2243.695202 &         dropout CCD row \\
 211909748 &  2317.909660 &  2317.950524 &        flagged cadences \\
 211936906 &  2336.012180 &  2336.093907 &        asteroid transit \\
 211955036 &  2338.463959 &  2338.484391 &               saturated \\
 211955036 &  2338.647844 &  2338.770434 &               saturated \\
 211972627 &              &              &            broken pixel \\
 211975426 &  2331.435440 &  2331.476304 &        asteroid transit \\
 212034371 &  2316.070632 &  2316.111496 &   no clear brightening  \\
 212034371 &  2338.361678 &  2338.422973 &           contamination \\
 212034371 &  2343.857791 &  2343.898654 &   no clear brightening  \\
 \hline
\end{tabular}

\tablefoot{Candidates detected in the time interval $[t_0,t_f]$. If $t_0,\,t_f$ are not given, the entire target was dropped.}
\end{table*}
\section{Synthetic Flare Injection and Recovery}
\label{appendix2}
\begin{figure}
\centering
\includegraphics[width=\hsize]{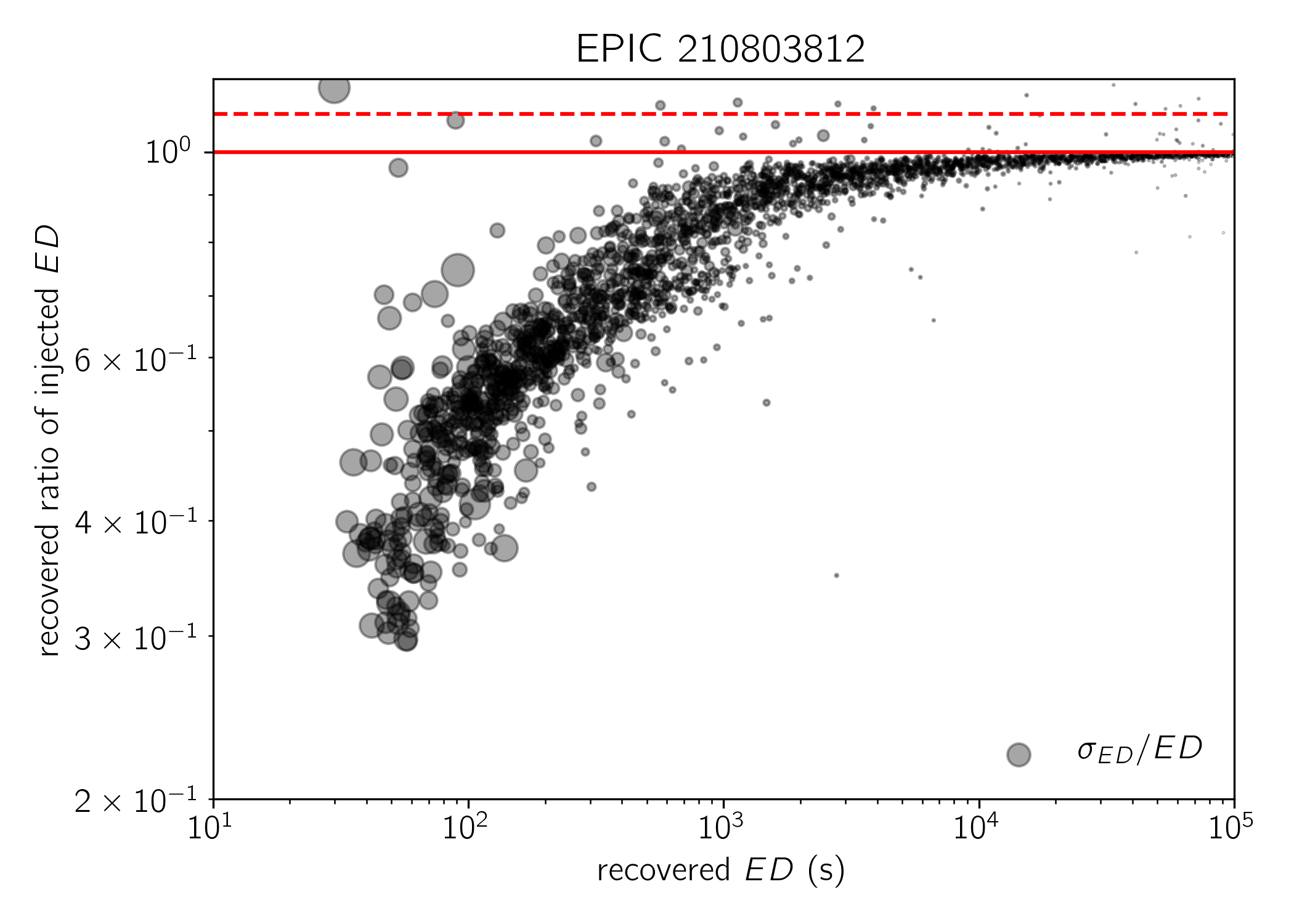}
\caption{Example of recovered energy ratios from synthetic flare injections. Ratio of the originally injected $ED$ recovered by \texttt{Appaloosa} in the \texttt{K2SC} de-trended long cadence LC for EPIC 210803812. The circle size represents the calculated relative uncertainty $\sigma_{ED}/ED$. The red lines are placed at $100\,\%$ (filled) and $110\,\%$ (dashed) energy recovery, given for orientation.}
\label{SFIenergyratio}
\end{figure}

\begin{figure}
\centering
\includegraphics[width=\hsize]{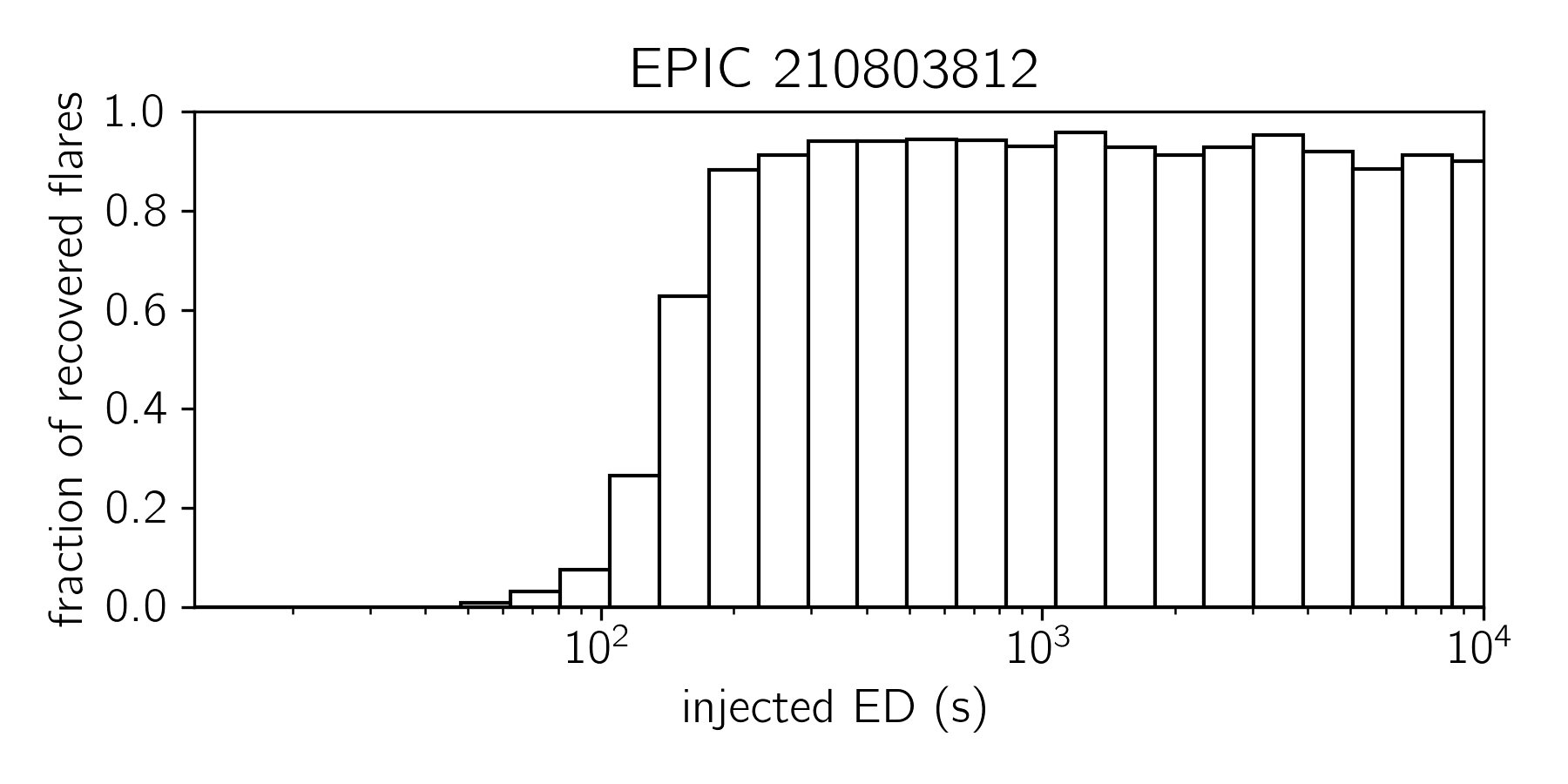}
\caption{Recovery rate of classic flare events as a function of $ED$ for EPIC 210803812.}
\label{SFIrecovery}
\end{figure}

\paragraph{Improvements to the Routine}
We carried out several enhancements to the original artificial flare injection subroutine to improve the validity of the returned recovery rates. The original code injected $100$ synthetic flares into a LC at once, a number too small to characterize the whole LC but too large to create mostly single and not superimposed events. We introduced an artificial flare grid in the amplitude-duration space and conducted more injections with fewer flares instead. We assumed that a LC's properties are uniform throughout the campaign during which it was recorded. The convergence of the recovery probability distribution was a reasonable expectation because the flare finding procedure is deterministic, i.e. the same fake flare infested LC yields the same recovered flares. Note that artificial flare injection cannot take care of false positive signals. We addressed the issue by manually vetting high-energy events, and comparing multiple LCs to identify field-wide systematics.
\newline
We could directly infer the recovery probability for the injected flare $ED$s. Observed flares' $ED$s, however, are systematically underestimated in data with low time sampling ($30$\,min) compared to $1$\,min cadence LCs, as \citet{2018ApJ...859...87Y} found for \textit{Kepler} data. We compared the injected with the recovered $ED$ for every artificial flare. From the ratio, i.e., the recovered share of the injected energy, we derived a correction factor that allowed us to ascribe both a more realistic flare energy and the corresponding recovery rate to each candidate~(see Fig. \ref{SFIenergyratio}). 
\newline
In summary, we obtained two $ED$-dependent corrections to the $ED$s and the recovered flare rates for each flare: an $ED$ correction factor and a correction to the Poissonian count uncertainty, respectively. We accounted for the systematic energy underestimation on a LC by LC basis~(Fig. \ref{SFIenergyratio}). Each LC obtained a unique flare recovery relation from the series of synthetic flare injections as a function of injected $ED$. This detection rate of synthetic events revealed the detection threshold for flares and provided a means to account for flares detected within the transition region by increasing the uncertainty on the detected number of flares accordingly~(Fig. \ref{SFIrecovery}).
\\
For M dwarfs, \citet{hawley_kepler_2014} found clear evidence for an exponential distribution of flare amplitudes as a function of duration. We enabled users of \texttt{Appaloosa} to generate artificial flares generously covering this empirical parameter space~(Fig. \ref{SFI}). The main advantage is a better coverage of the low-energy/low-detectability end of the expected flare distribution which is also most densely populated.
\paragraph{Realistic Synthetic Flare Injections}
Caution is recommended regarding the universality of injected flare shapes. Although the self-similarity of flares with different durations and amplitudes for a large parameter range is striking, the semi-analytical flare model used to generate synthetic events is derived from the flare distribution of a single star, GJ~1243~\citep{davenport_kepler_2014}, a bright dMe flare star in the northern sky~\citep{lepine_spectroscopic_2013}. Furthermore, \citet{davenport_kepler_2014} found a small portion ($1.3\,\%$) of flares on that star that could only poorly be fitted by this single star model. The injection procedure is therefore tailored to characterize a LC's quality with respect to the most common flares on GJ 1243. It can be extended to treat superpositions of classical flares, i.e. multiflare events, that can amount to $15\,\%$ of all events~\citep{davenport_kepler_2014} and are occasionally created, but complex shapes cannot be represented by the injections, although they can be recovered by the detection routine we employ here. We acknowledge that there is room for improvement in this respect.
\paragraph{Light Curve Quality from Synthetic Flare Injections}

\label{LCqualitySFI}

Relying on LCs alone for the detection of flare candidates requires detailed quality assessment of each LC. We therefore injected and recovered $20$ randomly distributed flares into every LC and repeated the process $300$ times each using \texttt{Appaloosa}. Of these injections $\sim5\,\%$, were dropped, because they were juxtaposed with more energetic events yielding more complex signatures. Consequently, $\sim 5,700$ synthetic flares injected into each LC $-$ a total of more than $9$\,million events to cover all LCs $-$ were used to retrieve information about their quality with respect to thresholds of detection, and systematic errors on energy recovery. For each cluster, this resulted in individual recovery rates based on the properties of the de-trended LCs, e.g. noise levels, time resolution, or variability time scales and magnitudes. Other properties may have been implicitly covered, because the procedure did not make any a priori assumptions about what affects flare signature detectability and to what degree. Overall, flare energies are systematically underestimated with significant uncertainties that become smaller with increasing energy~(see Fig. \ref{SFIenergyratio}). Recovery rates typically exceed $80\,\%$ as soon as some LC specific threshold for detection is exceeded, e.g. the time resolution limit. However, $100\,\%$ recovery of both events and their energies is never achieved due to the cutoff of continuous observation periods~(Fig. \ref{SFIrecovery}). 
\label{corrs}

\end{appendix}

\end{document}